\documentclass[a4paper,11pt]{article}

\usepackage{jheppub} 
\usepackage[T1]{fontenc} 
\usepackage[italian,english]{babel}
\usepackage{hyperref}
\usepackage{ifpdf}
\usepackage{subfigure}
\usepackage{tikz}
\usepackage[normalem]{ulem}
\usepackage{slashed}
\usepackage{pgffor}							
\usepackage{amssymb}
\usepackage{amsfonts}
\usepackage{epsf}
\usepackage{rotating}
\usepackage{graphicx}
\usepackage{amsmath}
\usepackage{fancyhdr}
\usepackage{lineno}
\usepackage{siunitx}
\usepackage{babel}
\usepackage{graphics}
\usepackage{pstricks}
\usepackage{color}
\usepackage{multirow}
\usepackage{float}
\usepackage{lineno}
\usepackage{mathtools}
\usepackage{stackengine}
\newcommand{\be}{\begin{equation}}
\newcommand{\ee}{\end{equation}}
\newcommand{\bea}{\begin{eqnarray}}
\newcommand{\eea}{\end{eqnarray}}

\newcommand{\aidm}{$A^{\prime}$iDM }
\usepackage{comment}

\author[a,b]{Abhishek Roy,}
\author[a,b]{Prasenjit Sanyal,}
\author[a,b]{Stefano Scopel,}
\affiliation[a]{Center for Quantum Spacetime, Sogang University, Seoul 121-742, South Korea}
\affiliation[b]{Department of Physics, Sogang University, Seoul 121-742, South Korea}
\emailAdd{abhishek@sogang.ac.kr}
\emailAdd{prasenjit.sanyal01@gmail.com}
\emailAdd{scopel@sogang.ac.kr}

\title{Dark Photon mediated Inelastic Dark Matter in Cosmology,  Astrophysics and Colliders}

\abstract{
We explore the phenomenology of the Dark Photon iDM ($A'\text{iDM}$) model, one of the simplest and most direct realizations of the inelastic Dark Matter (iDM) scenario. In this framework, the Standard Model (SM) is extended by a dark sector with an additional $U(1)_D$ gauge symmetry, with dark coupling $\alpha_D$. All SM particles are neutral under this symmetry, which couples to the SM hypercharge gauge boson via a kinetic mixing parameter $\epsilon$. Our work expands upon existing $A'\text{iDM}$ literature in three key ways: we move beyond specific benchmark points, restrict our focus to cosmologically viable configurations, and evaluate the complementarity between accelerator and astrophysical signals. The model features a dark photon $A'$ with mass $M_{A'}$ and two Majorana states, $\chi_1$ and $\chi_2$, with a mass splitting $\delta = M_{\chi_2} - M_{\chi_1} > 0$, where $\chi_1$ serves as the dark matter candidate. We perform a comprehensive scan of the parameter space ($M_{\chi_1}$, $\delta$, $M_{A'}$, $\epsilon$, $\alpha_D$) analyzing the $\chi_1$ relic abundance, direct and indirect detection prospects, and potential signals at accelerators and in astrophysics. Contrary to suggestions from previous studies restricted to specific benchmarks, our scan demonstrates that $\alpha_D = \alpha_{EM}$ is not phenomenologically disfavored. We also find that when the $\chi_1$ relic density matches observational data, direct and indirect searches become kinematically inaccessible. However, we show that the projected luminosity of FASER---a dedicated Long-Lived Particle (LLP) detector at the LHC---can probe or exclude the model's parameter space for $M_{\chi_1} \lesssim 7\text{ GeV}$, $100\text{ MeV} \lesssim \delta \lesssim 300\text{ MeV}$, and $M_{A'} \lesssim 25\text{ GeV}$. This reach could be significantly extended by the proposed FASER 2 upgrade for the High-Luminosity LHC. Interestingly, the parameter space accessible via accelerator LLP searches partially overlaps with the region probed by $\chi_1$ capture in neutron stars. This capture process is expected to heat nearby neutron stars to approximately $2000\text{ K}$, offering a promising, though challenging, signature for future infrared telescope observations.
}


\begin{document}
\preprint{CQUeST-2026-0771}

\maketitle

\section{Introduction}
\label{sec:introduction}

The search for Weakly Interacting Massive Particles (WIMPs),  the most natural candidates to provide the Cold Dark Matter that is supposed to have triggered galaxy formation and is believed to provide about 25\% of the density of the Universe under an invisible form~\cite{Planck:2018vyg}, has proved elusive so far. The increasingly stringent experimental limits put growing tension on the standard thermal decoupling scenario, which requires a minimal size for the WIMP couplings with SM particles in the early Universe in order to yield the correct value for the relic density. This motivates the interest toward scenarios where WIMPs interact with SM particles comparatively faster in the early Universe than in today DM searches. 

An example of such scenarios is inelastic dark matter (iDM), which was first proposed to explain the DAMA anomaly in the context of dark matter direct detection (DD) searches~\cite{inelastic_Tucker-Smith:2001}. In this class of models a Dark Matter (DM) particle $\chi_1$ of mass $M_{\chi_1}$ interacts with atomic nuclei exclusively by upscattering to a second heavier state $\chi_2$ with mass $M_{\chi_2}$ = $M_{\chi_1}$ + $\delta$. A peculiar feature of iDM is that there is a minimal WIMP incoming
speed in the target frame matching the kinematic threshold for inelastic upscattering events and given by:

\begin{equation}
    v_{T*}=\sqrt{\frac{2\delta}{\mu_{\chi T}}},
    \label{eq:vstar}
\end{equation}

\noindent with $\mu_{\chi T}$ the WIMP--target reduced mass. In particular, the WIMPs in our Galaxy are much slower than in the early Universe, so that such velocity-dependent threshold can suppress the signals in DD searches and at the same time allow for a large--enough WIMP annihilation rate at the time of the WIMP decoupling from the thermal bath to yield the observed relic abundance. Moreover, if at the present time all the $\chi_2$ heavy states have decayed away, the $\chi_1 \chi_2$ annihilation process in the halo of our Galaxy is not possible and the $\chi_1 \chi_1$ can be suppressed, explaining why also indirect searches have proved ineffective so far. 

In order to observe the iDM scenario in today's Universe the $\chi_1$ particles need to be accelerated beyond the kinematic threshold. The heavy state $\chi_2$ can be easily produced in accelerators, where the iDM kinematic threshold is negligible ~\cite{Izaguirre:2015zva, Berlin:2018jbm, Kang:2021oes, Lu:2023cet, Jodlowski:2019ycu, Bertuzzo:2022ozu, Liu:2025abt}. In such case, depending on the value of $\delta$, the decay of the $\chi_2$ particle can be observed at the Large Hadron Collider (LHC) through a displaced vertex either inside the the detector (for example in CMS, ATLAS and LHCb)  or in dedicated detectors positioned at some distance away from the interaction points (for example, FASER \cite{Feng:2017uoz, FASER:2018eoc, FASER:2018bac, FASER:2019aik, FASER:2022hcn}, MATHUSLA \cite{Curtin:2018mvb, MATHUSLA:2020uve, MATHUSLA:2025eth}, CODEX-b \cite{Gligorov:2017nwh, CODEX-b:2019jve, CODEX-b:2025rck}, AL3X \cite{Gligorov:2018vkc}, MAPP \cite{Pinfold:2019zwp}, ANUBIS \cite{ANUBIS:2025sgg} and FACET \cite{Cerci:2021nlb}). In particular, among the latter FASER 
is currently operating in the LHC Run 3 data taking period. In the following we will henceforth restrict our study to FASER and its proposed upgrade FASER~2 for the High-Luminosity phase of the LHC (HL-LHC).

Moreover, the velocity threshold induced by the mass splitting can be overcome when the $\chi_1$ particles are accelerated by the gravitational potential of a celestial body before up--scattering to the $\chi_2$ state off the nuclei inside it. If the DM particle is captured inside the star this can lead to detectable astrophysical signals~\cite{idm_wd_2010, higgino_wd_2019, idm_wd_2010_Hooper, improved_WD_2021, Biswas:2022cyh, IDM_neutron_stars_2022, improved_neutron_stars_2020, tanedo_2018, Bell_NS_2018, Tait_2022,Alvarez:2023fjj}. 

One of the simplest and more straightforward realization of the iDM scenario is in the Dark Photon model (hereafter referred to as \aidm). The dark photon is a portal between the SM and dark-sector states~\cite{Tucker-Smith:2001myb,Tucker-Smith:2004mxa,Berlin:2018jbm,Bertuzzo:2022ozu,Jodlowski:2019ycu} that represents a well-motivated framework to circumvent longstanding constraints on dark matter masses below the weak scale, such as the $Z$ invisible decay limits~\cite{Kearney:2016rng,Escudero:2016gzx} or the Lee-Weinberg bound~\cite{Lee:1977ua} (which excludes thermal relics below $\mathcal{O}(1)$ GeV). In the \aidm scenario the Standard Model is extended by a dark sector containing an additional $U(1)_D$ gauge symmetry (under which all SM particles are neutral and that couples to the SM hypercharge gauge boson through kinetic mixing), and a Dirac fermion which splits into the two Majorana states $\chi_1$ and $\chi_2$~\cite{Cui:2009xq}.

In spite of its simplicity and of the limited number of its free parameters~\cite{Chun_2010}, a systematic discussion of the phenomenology of \aidm is still lacking in the literature, were it has been only discussed in specific benchmarks~\cite{Izaguirre:2015zva,Berlin:2018jbm,Jodlowski:2019ycu,Liu:2025abt,Bertuzzo:2022ozu,Kang:2021oes,Lu:2023cet,Graham:2021ggy}. The goal of the present paper is to fill this gap by exploring in a systematic way a Dark Photon iDM scenario where the relic abundance, DD and indirect detection as well as potential signals from astrophysics and from accelerators are properly taken into account. In particular, we will show that fixing the $U(1)_D$ gauge coupling to the electromagnetic one is not phenomenologically disfavored, as one could conclude from analyses that can be found in the literature that are limited to specific benchmarks~\cite{Berlin:2018jbm,Graham:2021ggy,Izaguirre:2015zva,Izaguirre:2015yja}.

The plan of the paper is the following. We outline the \aidm scenario in Section~\ref{sec:Model}; in Section~\ref{sec:constraints} we discuss various existing constraints on the model (namely from colliders in Section~\ref{sec:collider_constraints} and from the relic density and BBN in Section~\ref{sec:early_universe_constraints}); in Section~\ref{sec:neutron_stars} we discuss the potential signal from Neutron Stars (NS); in Section~\ref{sec:Collider} we discuss how the model can be tested in accelerators. In particular, in Sections~\ref{sec:constraints}, \ref{sec:neutron_stars} and \ref{sec:Collider} we show  our quantitative results in a 3--dimensional subspace of the full parameter space (Scan 1 in Table~\ref{Table:Scan_parameter}). Finally, we provide our discussion and conclusions in Section~\ref{sec:discussion_conclusions}. In Appendix~\ref{app:5dim} we show that when the scan is extended to the full 5 dimensions (Scan 2 in Table~\ref{Table:Scan_parameter}) the results are qualitatively similar to those for Scan 1, while in Appendix~\ref{Appendix: collider} we provide additional details about the effect of the Lorentz boost on the $\chi_2$ decay length.

\section{The Model}\label{sec:Model}
We consider the simple extension of the standard model (SM) symmetry with an additional $U(1)_D$ gauge symmetry, whose associated gauge boson is referred to as the dark photon, $ A^\prime$. The dark photon couples to the SM hypercharge gauge boson through kinetic mixing, which can be expressed as~\cite{Holdom:1985ag,delAguila:1988jz},
\begin{align}
\mathcal{L}_{int}\supset\frac{\epsilon}{2\cos\theta_W}A^{\prime \mu\nu}B_{\mu\nu},
\label{eq:dark_photon_Kinetic_mixing}
\end{align}
where $A^{\prime\mu\nu}$ and $B^{\mu\nu}$ are the field strength tensors associated with the $U(1)_D$ and SM $U(1)_Y$ gauge groups, $\cos\theta_W$ is the cosine of the weak mixing angle and $\epsilon$ 
parametrizes the strength of the kinetic mixing, with $e_D$ and $e$ denoting the $U(1)_D$ gauge and electromagnetic couplings respectively. The kinetic mixing can be thought to arise radiatively from particles in the loop that are charged under both $U(1)$ groups. Furthermore, in our analysis we will consider a dark photon with non--vanishing mass $M_{A^{\prime}}$, but will remain agnostic about the mechanism of its generation. The most plausible scenario involves a scalar degree of freedom breaking the $U(1)_D$ symmetry spontaneously, thereby giving rise to the dark photon's mass while being decoupled from the theory. 
The dark photon interacts with SM fermions through,

\begin{align}
\mathcal{L}_{int}=\sum_f \left (C^V_{A^{\prime}ff}\bar{f}\gamma^{\mu}f+ C^A_{A^{\prime}ff} \bar{f}\gamma^{\mu}\gamma_5 f\right ).
\label{eq:dark_photon_fermion}
\end{align}
\noindent 

In particular, when $M_{A^\prime}\lesssim M_Z$ the axial coupling is negligible compared to the vector coupling, and the latter is given by\footnote{When $M_{A^{\prime}}\ll M_Z$ Eq.~(\ref{eq:dark_photon_fermion}) reduces to $\mathcal{L}_{int}=\epsilon e\sum_f Q_f \bar{f}\gamma^{\mu}f$. For our numerical analysis we use for $C^V_{A^{\prime}ff}$ the general expression presented in Ref.~\cite {Curtin:2014cca,Berlin:2018jbm}.}~\cite{Hoenig:2014dsa},
\begin{align}
C^V_{A^\prime\bar{f}f}\approx -\frac{\epsilon}{\cos\theta_W}\frac{M^2_Z\cos\theta_{W}eQ_{f}-M^2_{A^\prime}gY_{f}}{M^2_{Z}-M^2_{A^\prime}},
\label{eq:dark_photon_fermion_coupling}
\end{align}
where $Y_f$ corresponds to the hypercharge of the SM fermion.

In this work, we consider an inelastic dark matter (iDM) scenario in which a Dirac fermion $\chi$ charged under $U(1)_D$ and described by the following Lagrangian~\cite{Cui:2009xq}:
\begin{align}
\mathcal{L}_{int}\supset \bar{\chi}i\gamma^{\mu}(\partial_{\mu}+ig_{D}A^\prime_{\mu})\chi-M_{\chi}\bar{\chi}\chi-\frac{1}{2}M_{L}(\bar{\chi^c}P_L\chi+h.c)-\frac{1}{2}M_{R}(\bar{\chi^c}P_R\chi+h.c),
\label{eq:dark_matter_interaction}
\end{align}
splits into two nearly degenerate Majorana states $\chi_{1,2}$ via the Majorana mass terms $M_{L,R}$ with $M_{\chi}\gg M_{L,R}$. In the mass eigenstate basis  the \aidm  Lagrangian~(\ref{eq:dark_matter_interaction}) can then be expressed as:
\begin{align}
\mathcal{L}_{int}\supset &\frac{1}{2}\bar{\chi}_1i\gamma^{\mu}\partial_{\mu}\chi_1-\frac{1}{2}M_{\chi_1}\bar{\chi}_1\chi_1 \nonumber\\
&+\frac{1}{2}\bar{\chi}_2i\gamma^{\mu}\partial_{\mu}\chi_2-\frac{1}{2}M_{\chi_2}\bar{\chi}_2\chi_2 \nonumber\\
&+ig_DA^\prime_{\mu}\bar{\chi}_1\gamma^{\mu}\chi_2+\frac{1}{2}g_{D}\frac{M_L-M_R}{2M_\chi}A^\prime_{\mu}(\bar{\chi}_2\gamma^\mu\gamma^5\chi_2-\bar{\chi}_1\gamma^\mu\gamma^5\chi_1)\nonumber\\
&+\mathcal{O}\left(\frac{M^2}{M^2_{\chi}}\right),
\label{eq:dark_matter_mass_eigen}
\end{align}
with $M_{\chi_{2,1}} = M_\chi \pm \frac{1}{2}(M_L + M_R)$ and $\delta \equiv M_{\chi_2} - M_{\chi_1} = M_L + M_R$ and where the lighter state $\chi_1$ is odd under the $\mathcal{Z}_2$ symmetry and a cosmologically stable DM candidate. In our work we impose $M_L=M_R$, under which the diagonal   interactions $A^\prime_{\mu}\chi_1\chi_1$ and $A^\prime_{\mu}\chi_2\chi_2$ vanishes identically\footnote{In the scenario $M_L\neq M_R$,  both diagonal and off-diagonal interactions arise, enriching the phenomenology~\cite{dallavalle_2024}.}. 
Hence, the phenomenology is determined by the following 5 parameters:
\begin{align}
    M_{A^\prime},\,M_{\chi_1},\,\delta,\,\epsilon,\,\alpha_{D}(=g^2_D/4\pi).
    \label{eq:free_parameters}
\end{align}
\noindent Both the annihilation cross section driving the relic abundance and the WIMP--nucleon cross section driving direct detection and capture in celestial bodies are driven by the 6--dimensional operator $\bar{f}\gamma^{\mu}f \bar{\chi_1}\gamma_{\mu}\chi_2$ mediated by and $A^{\prime}$ propagator, and are proportional to the combination $\zeta=\epsilon^2\alpha_D$ (see Eqs.~(\ref{eqn:DM_SigmaV}) and (\ref{eq:dd_coupling})). In our analysis we will perform two scans. In the first (Scan 1), whose numerical results will be used in Sections~\ref{sec:constraints}, \ref{sec:neutron_stars} and \ref{sec:Collider}, in order to reach an optimal density of points and to simplify visualization we will set $\alpha_D=\alpha_{EM}$ and $\epsilon=\epsilon_{\max}$ ($\zeta = 7.3\times 10^{-6}$). In the second (Scan 2) we will allow both $\alpha_D$ and $\epsilon$ to vary in the ranges $\alpha_{EM}\le \alpha_D\le$ 0.5 and $\epsilon_{\max}/10\le\epsilon\le\epsilon_{max}$ ($5\times 10^{-4}\le\zeta\le 7.3\times 10^{-7}$). The numerical results of Scan 2 will be shortly discussed in Appendix~\ref{app:5dim}, since they are qualitatively similar to those of Scan 1. 
Moreover, we set $M_{A^\prime}\le$ 60 GeV in order to avoid the parameter space where $M_{A^{\prime}}\rightarrow M_Z$ and the kinetic mixing parameter $\epsilon$ is strongly constrained by precision studies at LEP~\cite{Hook:2010tw,Curtin:2014cca} and displaced muon jet searches (DMJ) at CMS~\cite{CMS:2023bay}. 
We also avoid in our scan the parameter space for which $M_{\chi_1}+M_{\chi_2}\ge M_{A^{\prime}}$ where the dark photon only decays in visible particles and $\epsilon$ is strongly constrained by accelerator bounds~\cite{Graham:2021ggy,Fabbrichesi:2020wbt}.

\begin{table}[h!]
  \centering
  \begin{tabular}{|c|c|c|}
    \hline
    Parameter & Scan 1  & Scan 2 \\
    \hline
    $M_{\chi_1}\ [\text{GeV}]$ & $[1\ ,\ 30]$ & $[1\ ,\ 30]$ \\
    $\delta\ [\text{GeV}]$ & $[10^{-4}\ ,\ 20]$ & $[10^{-4}\ ,\ 20]$ \\
    $M_{A'}\ [\text{GeV}]$ & $[10\ ,\ 60]$ & $[10\ ,\ 60]$ \\
    $\epsilon$ & $\epsilon_{\text{max}}(M_{A'})$ & $[10^{-3},\epsilon_{\text{max}}(M_{A'})]$ \\
    $\alpha_D$ & $\alpha_{EM}$ & $[\alpha_{EM},0.5]$ \\
    \hline
  \end{tabular}
  \caption{Input parameters used in the two numerical scans of our analysis. $\epsilon_{\text{max}}(M_{A'})$ is the maximum value of $\epsilon$ allowed by the LEP for a given $M_{A^\prime}$~\cite{Hook:2010tw,Curtin:2014cca}, with $M_{A^\prime}>M_{\chi_1}+M_{\chi_2}$ condition. }
	\label{Table:Scan_parameter}
\end{table}

\section{Existing Constraints}
\label{sec:constraints}
In this Section we provide a list of the constraints from existing collider and DM observable bounds that we apply to the parameter space of the  \aidm scenario. In particular, for our phenomenological analysis we perform a flat random scan over the free parameters of Eq.~(\ref{eq:free_parameters}) within the ranges indicated in Table~\ref{Table:Scan_parameter}. As already pointed out the upper bound $M_{A^{\prime}}\le$ 60 GeV is chosen to avoid the peculiar regime $M_{A^{\prime}}\rightarrow M_Z$ of maximal mixing between the dark photon and the SM $Z$ boson. 

\subsection{Collider Constraints}
\label{sec:collider_constraints}
We consider the bounds arising from electroweak precision measurements performed at LEP, which excludes $\epsilon\gtrsim \mathrm{few}\times10^{-2}$ for dark photon masses $M_{A^\prime}\gtrsim 10\,\rm{GeV}$~\cite{Hook:2010tw,Curtin:2014cca}. The constraint is strongest when the mass of the dark photon is close to that of the $Z$ boson, and becomes progressively weaker for $M_{A^\prime}<M_Z$. For Scan 1 we fix $\epsilon$ to the maximum value allowed by LEP electroweak precision constraints for a given $M_{A'}$, i.e. $\epsilon = \epsilon_{\rm{max}}(M_{A^\prime})$. Specifically, for $M_{A'} \lesssim 10\,\mathrm{GeV}$, monophoton searches at BaBar provide more stringent constraints than LEP, excluding $\epsilon \gtrsim 10^{-3}$ from the process $e^+e^- \to \gamma A' \to \gamma \chi_1 \chi_2$ using an integrated luminosity of $53\,\mathrm{fb}^{-1}$ collected at the $\Upsilon$ resonance
~\cite{BaBar:2017tiz,BaBar:2008aby}. For $M_{A^\prime}<1\,\mathrm{GeV}$, $\epsilon$ is subject to stringent constraints from several experiments, such as LSND~\cite{deNiverville:2011it,LSND:2001akn}, E137~\cite{Bjorken:1988as,Batell:2014mga}, NA64~\cite{na64_2025} and others. We refer the readers to Refs.~\cite{Essig:2013lka,Graham:2021ggy,Berlin:2018pwi,Izaguirre:2017bqb, ship_2025} for comprehensive reviews of the limits on $\epsilon$ across different ranges of $M_{A^\prime}$. We therefore restrict our analysis to $M_{A'} \geq 10\,\mathrm{GeV}$ to avoid the low-mass regime with stringent bounds on $\epsilon$, and consider only constraints from LEP, taking $\epsilon_{\rm{max}}(M_{A^\prime})$ from the corresponding solid line in Fig.~1 of Ref.~\cite{Berlin:2018jbm}.

Since $M_{\chi_2}\gtrsim M_{\chi_1}$, the heavier state $\chi_2$ decays into $\chi_1\bar{f}f$ mediated by an off-shell $A^\prime$. The heavier state $\chi_2$ can be long-lived for a highly compressed mass spectrum ($\delta \ll 1$), allowing it to decay outside the LHC detector volume, thereby leading to a monojet signature. In this case, the process $pp\to A^\prime/Z \to j + \chi_2 \chi_1$ produces a hard jet recoiling against the invisible $\chi_2 \chi_1$ pair escaping the detector. We find that in the parameter space of our scenario the production cross-section for this process is suppressed due to the high $p_T(j)\geq200\,\rm{GeV}$ cut and is one order of magnitude smaller than the ATLAS sensitivity~\cite{ATLAS:2021kxv}\footnote{To calculate the relevant monojet cross section we  implement the Dark Photon model in \texttt{Feynrules}~\cite{Alloul:2013bka}, generate the \texttt{CalcHEP} file~\cite{Belyaev:2012qa} and feed it to \texttt{micrOMEGAs}~\cite{Alguero:2023zol,Barducci:2016pcb}.}.

The decay of the $\chi_2$ particle has also been searched for inside LHC detectors. CMS~\cite{CMS:2023bay} published some bounds searching for displaced muons, missing transverse energy and a jet, albeit only in some specific benchmark points. Recasting such bounds in the general parameter space of \aidm would require a dedicated full simulation. On general grounds, such searches are sensitive to large values of the $\delta$ parameter, due to the short required decay length. On the other hand astrophysical signals are more sensitive to low values of $\delta$ (see Section~\ref{sec:DM_NS}), corresponding to configurations where the distance between the interaction point and the detector is much larger, as for the FASER detector (see Section~\ref{sec:Collider}). For this reason, and with the goal to focus on the prospects of probing \aidm both through astrophysical observations and accelerator experiments, we postpone a full discussion of DMJ searches for future work. 

Additionally, we also enforce the partial decay width of $Z\to\chi_1\chi_2$ to satisfy $\Gamma^{\rm inv}_Z\lesssim\mathcal{O}(10^{-3})\,\rm{GeV}$~\cite{ParticleDataGroup:2024cfk}, as $A^\prime$-$Z$ mixing can modify the $Z$ decay width for $M_{A^\prime}>10\,\rm{GeV}$.
 
\begin{figure}[]
	\centering
	\includegraphics[height=6.cm,width=10.cm]{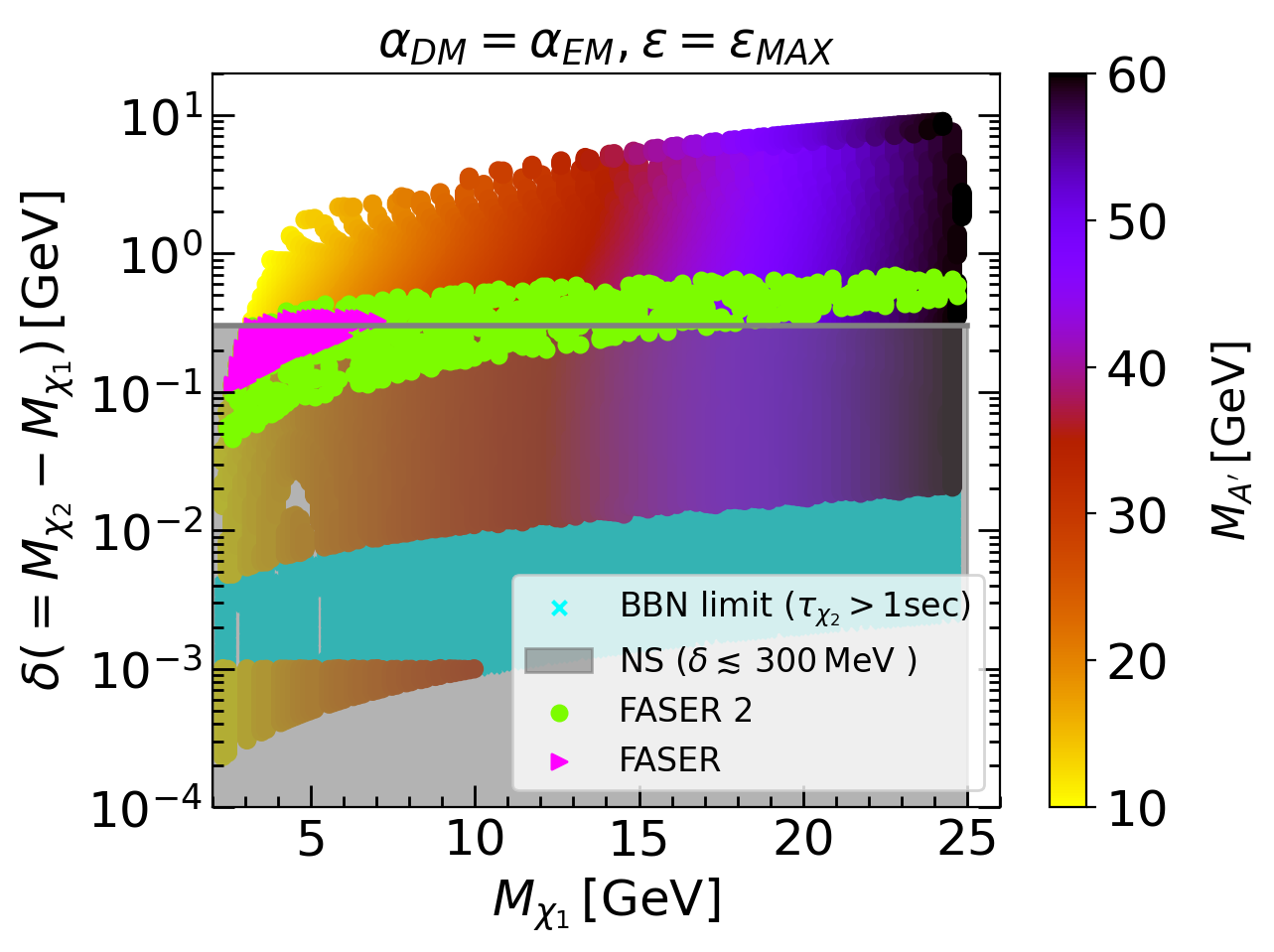}
	\caption{Scatter plots in $M_{\chi_1}-\delta$ plane with color pallets corresponding to $M_{A^\prime}$ (Scan 1, see Table~\ref{Table:Scan_parameter}). Note that all the points satisfy collider and DM observables constraints except the cyan colored points, which are disallowed by BBN. The magenta and green areas can be probed by Long Lived Particle searches at FASER and FASER~2  (see Section~\ref{sec:Collider}). }
	\label{fig:DM_Relic_BBN}
\end{figure}
\subsection{Dark Matter Observables Constraints}
\label{sec:early_universe_constraints}
As mentioned in Section~\ref{sec:Model}, in our model $\chi_1$ serves as viable WIMP DM candidate. In the early Universe, $\chi_1$ thermalizes with the thermal bath, and its number density is diluted primarily through the co-annihilation process, i.e., $\chi_1\chi_2\to A^\prime,Z\to \bar{f}f$ and subdominantly by $\chi_1\chi_1\to\chi_2\chi_2$ (via $A'$ exchange), which is phase-space suppressed. The evolution of the number density of DM $n_{\chi_1}$ is governed by the Boltzmann equation~\cite{Gondolo:1990dk,Kolb:1990vq}, 
\begin{align}
    \frac{dn_{\chi_1}}{dt}+3Hn_{\chi_1}=-\langle\sigma v\rangle_{eff}\left(n^{eq\,2}_{\chi_{1}}-n^{2}_{\chi_{1}}\right),
    \label{eqn:B_eqn}
\end{align}
where $n^{eq}_{\chi_{1}}$ denotes the equilibrium number density of $\chi_1$, $H$ is the Hubble expansion rate, and $\langle\sigma v\rangle_{eff}$ represents the DM thermal average effective cross-section, which can be approximated by~\cite{Berlin:2018jbm},
\begin{align}
    \langle\sigma v\rangle_{eff} \sim \mathcal{O}(10^{-2})\left ( \frac{\alpha_D}{1/137}\right )\frac{\epsilon_{max}^2(M_{A^\prime})M^2_{\chi_1}}{M^{4}_{A^\prime}}e^{-\delta/{T}}\,,
    \label{eqn:DM_SigmaV}
\end{align}
where $T$ is the temperature of the thermal bath. It is important to highlight that $\chi_2$ remains in thermal equilibrium as long as the annihilation process $\chi_2 \chi_2 \to \chi_1 \chi_1$ and the decay $\chi_2 \to \chi_1 \bar{f} f$ do not freeze out. For $\delta/T \ll 1$, $\chi_1$ and $\chi_2$ co-annihilate efficiently and the relic abundance is fixed by their simultaneous freeze out. In contrast, for $\delta/T \gg 1$ the $\chi_2$ state decouples earlier, leaving $\chi_1$ without an efficient co-annihilating partner and resulting in a DM relic abundance that exceeds the observed value. To calculate the relic abundance we solve Eq.~(\ref{eqn:B_eqn}) numerically using \texttt{micrOMEGAs}~\cite{Alguero:2023zol}, which includes all the relevant annihilation and co-annihilation processes. We then impose that the DM relic density falls within the $2\sigma$ PLANCK interval~\cite{Planck:2018vyg},
\begin{align}
    \Omega^{\rm{Obs}}_{\chi_1}h^2=0.12\pm0.0012.
    \label{eqn:obs_dark_matter}
\end{align}
%
%

In particular, in Figure~\ref{fig:DM_Relic_BBN}, we show a scatter plot in the $M_{\chi_1}-\delta$ plane of the configurations of Set 1 of our scenario whose relic abundance complies with Eq.~(\ref{eqn:obs_dark_matter}). The color pallets correspond to the value of $M_{A^\prime}$. Since the $\chi_1$ decouples from the thermal bath when the temperature drops to approximately $T \simeq M_{\chi_1}/25$ it follows from Eq.~(\ref{eqn:DM_SigmaV}) that co-annihilation is the dominant process when $\delta \lesssim 0.04\,M_{\chi_1}$ (corresponding to $\delta/T\simeq 1$ in the exponential). Therefore, we can see that for a given $M_{\chi_1}$ the upper limit on $\delta$ increases with $M_{\chi_1}$. Furthermore, for a given $\delta$, $M_{A'}$ increases with increasing $M_{\chi_1}$ to compensate for the scaling $M_{\chi_1}^2 / M_{A'}^4$ in Eq.~(\ref{eqn:DM_SigmaV}). 
Notice that in Fig.~\ref{fig:DM_Relic_BBN} the points extend up to $M_{\chi_1}\lesssim$ 25 GeV because of the $M_{A^{\prime}}\lesssim$ 60 GeV upper cut in our scan.

Since $\delta/T\ll 1$ from the requirement of the DM relic abundance, $\chi_2$ can be long-lived due to a compressed mass spectrum between $\chi_1$ and $\chi_2$. Nevertheless, $\chi_2$ decays well before the recombination epoch, thus resulting in the co-annihilation process being strongly suppressed at the time of recombination. In addition, the process $\chi_1\chi_1\rightarrow A^{\prime *}A^{\prime*}\rightarrow f\bar{f} f^{\prime}\bar{f}^{\prime}$ (mediated by a t-channel $\chi_2$) is velocity suppressed.  Consequently, residual annihilation of $\chi_1$, either through co-annihilation or pair annihilation, is inefficient at injecting energy into the primordial plasma. Therefore, CMB observations~\cite{Planck:2018vyg, CarrilloGonzalez:2021lxm} do not impose any constraint on our scenario. Note that for $\delta>1\,\rm{MeV}$ and and $\tau_{\chi_2} \gtrsim 1.0\,\mathrm{sec}$, $\chi_2$ decays primarily to charged leptons and hadrons; these decay products can induce photodissociation and hadro-dissociation of light elements and thereby alter the standard big bang nucleosynthesis (BBN) prediction~\cite{Kawasaki:2017bqm}. As a consequence, to avoid conflict with BBN, we impose that the lifetime of $\chi_2$ satisfies $\tau_{\chi_2} \lesssim 1.0\,\mathrm{secs}$ for  $\delta>1\,\rm{MeV}$. On the contrary, for $\delta<1\,\rm{MeV}$ when $\chi_2$ has a lifetime $\tau_{\chi_2}\gtrsim1.0,\mathrm{sec}$ it undergoes the three-body decay $\chi_2\rightarrow\chi_1\nu_{\rm inj}\bar{\nu}_{\rm inj}$. These injected neutrinos $\nu_{\rm inj}$ are inefficient at leaking energy into electromagnetic energy, either through co-annihilation, $\nu_{\rm inj}+\bar{\nu}_{bg}\to l^+l^-$, or pair annihilation, $\nu_{\rm inj}+\bar{\nu}_{\rm inj} \to l^+l^-$, where $\nu_{bg}$ denotes the background neutrinos~\cite{Kanzaki:2007pd,Li:2025clq}. Note that the co-annihilation process requires neutrino energies of order $m_e^2/T_\gamma\simeq {\cal O}(10^2)$ MeV, while pair annihilation requires energies of order $m_e$ to overcome the relevant energy threshold, with $T_\gamma$ the background photon temperature~\cite{Leng:2026iqm}. For $\delta<1\,\rm{MeV}$, the injected neutrinos do not have sufficient energy to efficiently initiate these processes, further suppressing the transfer of energy into the electromagnetic plasma. Hence, they are also not energetic enough to efficiently initiate photodissociation processes and are also inefficient in producing observable CMB spectral distortions. We have also explicitly calculated the contribution of the injected neutrinos to the effective number of relativistic species, $N_{\rm eff}$, and verified that it remains negligible, with $\Delta N_{\rm eff}\ll0.1$. Therefore, while an interval 1 MeV $\lesssim\delta\lesssim$ a few MeV (depending on the $\chi_1$ mass) is excluded by BBN, a region in the parameter space allowed by BBN, CMB spectral-distortion, and $N_{\rm eff}$ constraints reappears for $\delta<1\,\rm{MeV}$ and $\tau_{\chi_2}\gtrsim1.0,\mathrm{sec}$. This results in the peculiar disconnected triangular shaped region in the lower left of Fig.~\ref{fig:DM_Relic_BBN}.

The values of $\delta$ that can be probed by direct detection (DD) experiments 
correspond to configurations for which 
the maximal speed of the WIMPs at the Earth's surface exceeds the kinematic threshold of Eq.~(\ref{eq:vstar}), $u_{max}>v_{T*}$, with $u_{max}\simeq$ 800 km/s obtained by combining the WIMP escape velocity in the Galaxy $u_{esc}\simeq$550 km/s ~\cite{vesc_Smith2006, vesc_Piffl2013} and the motion of the solar system, $v_0\simeq$ 220 km/s~\cite{SHM_maxwell_Green2011}. In particular, in Fig.~\ref{fig:DM_Relic_BBN} all the configurations at low $\delta$ correspond to a light $m_{\chi_1}$ ($m_{\chi_1}\lesssim$ a few GeV), where $u_{max}<v_{T*}$ for all the nuclear targets used in DD. 
On the other hand the elastic process is always negligible, because the relevant loop-mediated cross section is suppressed by the eighth power of $M_{A'}$~\cite{Berlin:2018jbm} and well below the neutrino floor never exceeding $10^{-49}\,\mathrm{cm}^2$. We conclude that our model does not produce any signal in DD experiments. 
Also indirect DM searches, that look for the decay products of DM annihilation in the halo of our Galaxy, cannot detect the \aidm scenario. In fact, the $\chi_1\chi_2$ annihilation process is not available, since at the present time all the $\chi_2$ particles have decayed away. Moreover, the four--fermion annihilation cross section for the process $\chi_1\chi_1\rightarrow A^{\prime *}A^{\prime*}\rightarrow f\bar{f} f^{\prime}\bar{f}^{\prime}$  is velocity suppressed ($p$--wave) and negligible~\cite{four_fermions_2018} at the WIMP speeds in the DM Halo ($\lesssim 10^{-3}c$). 


\section{Dark Matter capture in Neutron Stars}\label{sec:DM_NS}
\label{sec:neutron_stars}

The capture process takes place when a $\chi_1$ particle in the dark halo of our Galaxy crosses a celestial body and, by the same WIMP–nucleus scattering process driving direct detection, loses enough of its energy to became gravitationally captured~\cite{Gould_Sun_1987, Gould_Earth_1987}.  This can lead to measurable effects, such as a flux of high–energy neutrinos~\cite{Silk_Olive_1985, capture_gould_1987}, or an increase in the temperature/luminosity of the celestial body~\cite{Goldman_Nussinov_1989}. In the case of inelastic scattering capture extends the kinematic reach of direct detection to larger values of the mass splitting $\delta$ when the gravitational acceleration of the celestial body drives the $\chi_1$ particles to higher speeds compared to what happens in terrestrial detectors.  In particular, if $u$ and $w$ are the speeds of the WIMP far away from the celestial body and at the target position $r$ inside it, one has
\begin{equation}
    w(r)^2=u^2+v^2_{\text{esc}}(r),
\end{equation}
\noindent with $v_{\text{esc}}(r)$ the escape velocity in the celestial body at $r$.  This implies that one can have $w\gg u_{max}$, depending on the intensity of the gravitational field. Indeed, at the Earth's surface the escape velocity is $\simeq$ 11 km/s and  $w\simeq u$, while that at the center of the Sun $w$ reaches 1600 km/s, allowing to extend the kinematically accessible values of $\delta$ to $\simeq$ 600 keV~\cite{idm_sun_Catena_2018}. In the case of compact stellar remnants the speed of the scattering WIMP can reach $\simeq$ a few $10^4$ km/s for a White Dwarf~\cite{improved_WD_2021} or a few $10^5$ km/s for a Neutron Star~\cite{improved_neutron_stars_2020}, potentially extending further the values of $\delta$ that can be probed up to several tens of MeV for the former and hundreds of MeV for the latter. Stellar remnants have the additional advantage that they can be cold enough to potentially detect the anomalous heating produced by the kinetic energy transfer in the scattering process, in absence of the $\chi_2\chi_1$ coannihilation process and with $\chi_1\chi_1$ annihilation strongly suppressed. In the following we will discuss the potential signal from $\chi_1$ capture in a neutron star close to the solar system with representative mass $M=M_*$ = 1.5 $M_\odot$ and radius $R=R_*$ = 10 km. The gravitational acceleration of such system drives the WIMPs to a maximal speed $w_{max}\simeq 0.8 c$ corresponding to $\delta\lesssim$ 300 MeV from the kinematic condition~(\ref{eq:vstar}).

The WIMP--nucleon inelastic scattering cross section that drives the capture process is given by~\cite{Bell_NS_2018}:

\begin{eqnarray}
    \sigma_{\text{inel}}&\simeq& \sigma_{\text{el}}\beta_{\text{cl}},\nonumber\\
    \beta_{\text{cl}}&=&\sqrt{1-\frac{4k\mu_{+}}{w^2}},
    \label{eq:sigma_inel}
\end{eqnarray}

\noindent with $\sigma_{\text{el}}$ the corresponding elastic cross section, that, unless  $M_{A^{\prime}}\rightarrow M_Z$, is driven by the vector coupling $C^V_{A^{\prime \bar{f}f}}$. In Eq.~(\ref{eq:sigma_inel}) $k=\delta/M_{\chi_1}$, $\mu_{+}=(1+\mu)/2$, $\mu=M_{\chi_1}/m_N$ and $m_N$ is the nucleon mass. The speed of the $\chi_1$ particles inside the neutron star requires to use the relativistic form for the elastic cross section~\cite{Bell_NS_2018}: 
\begin{figure}[]
	\centering
	\includegraphics[height=6.cm,width=10.cm]{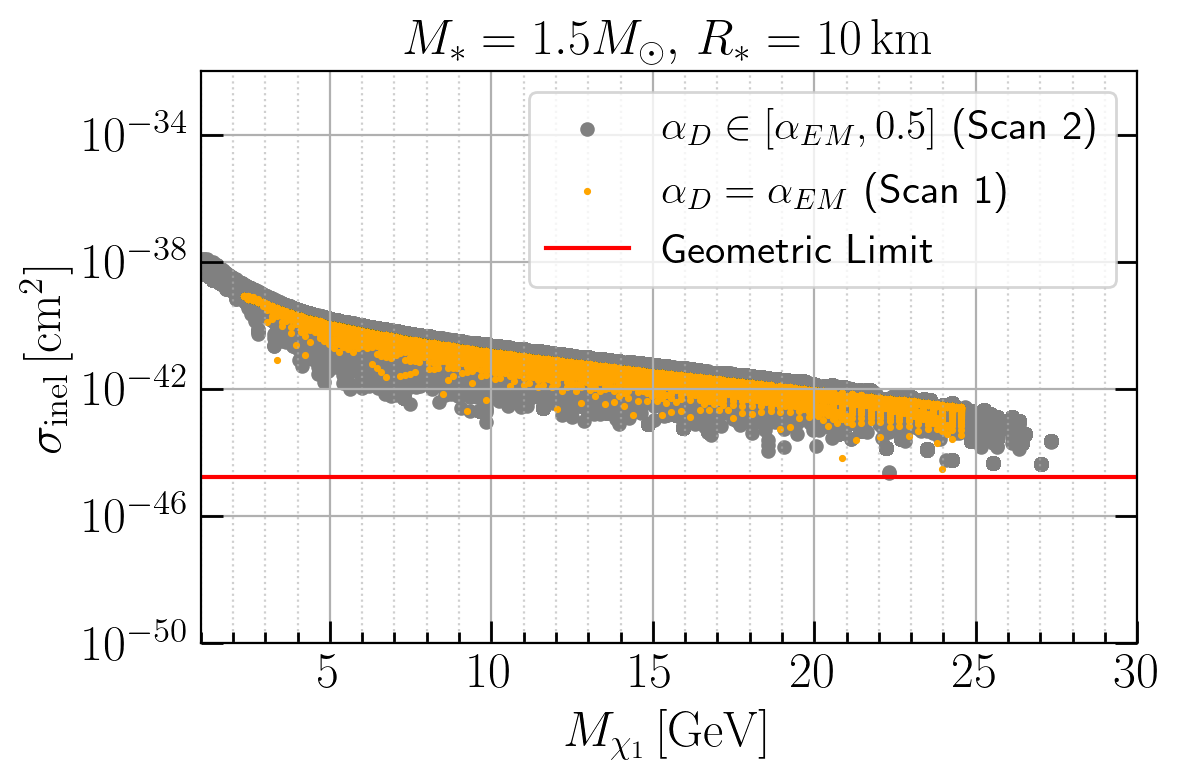}
	\caption{Scatter  in the $(M_{\chi_1},\,\sigma_{\text{inel}})$ plane showing points consistent with relic density, and known experimental constraints (Scan 1, see Table~\ref{Table:Scan_parameter}). The horizontal line indicates the geometric limit $\sigma_{\text{geom}}$ for $M=M_*$ and $R=R_*$; all satisfy $\sigma_{\text{inel}}>\sigma_{\text{geom}}$, implying geometric capture saturation.}
		\label{fig:DM_NS}
\end{figure}

\begin{equation}
    \frac{d\sigma_{\text{el}}}{d\cos\theta}(s,t)=\frac{1}{\Lambda^4} \frac{2(\mu^2+1)^2 M_{\chi_1}^4 - 4(\mu^2+1)\mu^2 s M_{\chi_1}^2 + \mu^4 (2s^2 + 2st + t^2)}{16\pi \mu^4 s},
\end{equation}

\noindent with, for a neutron target:

\begin{equation}
    \frac{1}{\Lambda^2}=\frac{g_D}{M_{A^{\prime}}^2} \left ( C^V_{A^{\prime}\bar{u}u}+2 C^V_{A^{\prime}\bar{d}d}\right ).
\label{eq:dd_coupling}
\end{equation}

\noindent Moreover: 

\begin{eqnarray}
    &&\frac{d \cos \theta}{dt} =\frac{B\mu^2 + 2\sqrt{B}\mu + B}{M_{\chi_1}^2 \delta t}, \nonumber \\
   &&t_{\min} = \frac{M_{\chi_1}^2 \left( B(k^2 + 2k + 2) - \delta t + \sqrt{B}k(k+2)\mu - 2 \right)}{B\mu^2 + 2\sqrt{B}\mu + B}, \nonumber \\
   &&t_{\max} = \frac{M_{\chi_1}^2 \left( B(k^2 + 2k + 2) + \delta t + \sqrt{B}k(k+2)\mu - 2 \right)}{B\mu^2 + 2\sqrt{B}\mu + B},\nonumber \\
   &&\delta t = \sqrt{1-B} \sqrt{B \left( k^4 \mu^2 + 4k^3 \mu^2 + 4k^2 (\mu^2 - 1) - 8k - 4 \right) - 4\sqrt{B} k(k+2) \mu + 4},
\end{eqnarray}
\noindent 
\noindent with $\theta$ the scattering angle in the center of mass, $s$ and $t$ the Mandelstam variables,

\begin{equation}
    B(M,r)=1-\frac{2GM}{c^2 r},
\end{equation}
\noindent and $G$ the gravitational constant. 

When the sum of the cross sections of all the neutron targets in the star exceeds the geometrical cross section:

\begin{equation}
    \sigma_{\text{geom}}=\frac{\pi R^2 m_n}{M},
    \label{eq:sigma_geom}
\end{equation}
\noindent the capture process is optically thick, so that all the $\chi_1$ particles along the star's path (including the focusing effect of its gravitational attraction) are captured and loose almost all their initial kinetic energy in the celestial body through a large number of scattering processes. In this case, assuming for the WIMPs a Maxwellian distribution with velocity dispersion $v_{\text{rms}}$, the capture rate $C$ (in $s^{-1}$) saturates its geometrical limit:

\begin{equation}
    C_{\star} = \frac{\pi R_{\star}^{2} (1 - B(M,R))}{v_{\star} B(M,R)} \frac{\rho_{\chi_1}}{m_{\chi}} \text{Erf} \left( \sqrt{\frac{3}{2}} \frac{v_{\text{rms}}}{v_{d}} \right)
    \label{eq:c_geom}
\end{equation}
\noindent with $v_{d}$ = $v_0$ if the neutron star is in the solar system and $\rho_{\chi_1}$ the DM density in the neighborhood of the Sun. 

In Fig~\ref{fig:DM_NS} $\sigma_{\text{inel}}$ is plotted versus $M_{\chi_1}$ both in Scan 1 of Table~\ref{Table:Scan_parameter}(yellow points) and in Scan 2 of the same Table (grey points) over the full set of configurations that yield the correct relic density and are allowed by the BBN limit. In the same plot the horizontal solid line represents the geometrical cross section $\sigma_{\text{geom}}$ for our benchmark neutron star with $M$ = $M_*$ and $R$ = $R_*$. Such plot shows that $\sigma_{\text{inel}}>\sigma_{\text{geom}}$ for all the allowed configurations of the model (irrespective on which scan is used), so that capture always saturates the geometric value of Eq.~(\ref{eq:c_geom}) when it is kinematically possible. This is due to the strong correlation between $\sigma_{\text{inel}}$ and $\langle\sigma v\rangle_{eff}$, that both depend on the combination $\alpha_D\epsilon^2/M_{A'}^4$. In other words, the minimal interaction strength required to avoid overclosure in the relic abundance ensures that capture in the neutron star always takes place in the optically--thick regime. As a consequence, the expected kinetic heating in Kelvin of the neutron star is given by the simple expression~\cite{Bramante_2017,tanedo_2018}:

\begin{eqnarray}
    T_{\text{kin}}&\simeq&\left[ \frac{\rho_{\chi_1} (1 - B) B}{4 \sigma_{SB} v_{\text{rms}}} \left( \frac{1}{\sqrt{B}} - 1 \right) \text{Erf} \left( \sqrt{\frac{3}{2}} \frac{v_{\text{rms}}}{v_{d}} \right) \right]^{1/4} \nonumber\\
    &=& 1700 \text{K} \left (\frac{\rho_{\chi_1}}{0.4\text{GeV cm}^{-3}} \right )^{1/4} F\left(\frac{v_{\text{rms}}}{230\text{km s}^{-1}}\right), 
    \label{eq:ns_heating}
\end{eqnarray}
\noindent where $\sigma_{SB}$ is the Stefan-Boltzmann constant,  $F(x)=[\text{erf}(x)/(x \text{erf}(1))]^{1/4}$ and, in the second line, $B=B(M_*,R_*)$ = 0.55,  $\rho_{\chi_1}\simeq 0.4\,\mathrm{GeV/cm^3}$ and $v_{\text{rms}}$ = $\sqrt{3/2} v_0$ $\simeq 270$ km/s (assuming hydrostatic equilibrium in the halo between pressure and gravity~\cite{Lewin_Smith_1996}). Notice that, provided that the optically thick limit is saturated (i.e. for cross sections exceeding the geometrical limit of Eq.~(\ref{eq:sigma_geom}) ) the expected amount of heating is quite insensible to the \aidm parameters, including the $\chi_1$ mass. In an ideal situation a neutron star as old as $2\times 10^{7}$ years~\cite{Yakovlev_2004} could potentially cool down to temperatures below Eq.~(\ref{eq:ns_heating}), although seizable background heating mechanisms have been discussed in the literature that could prevent this to happen~\cite{raj_2024, Vortex_creep_heating_2024} . If in future observations with infrared telescopes such a source could indeed be detected this would in principle allow to probe the full range of configurations of Figs.~\ref{fig:DM_Relic_BBN} and~\ref{fig:DM_NS} using the DM heating effect. In Fig.~\ref{fig:DM_Relic_BBN} the region accessible to capture in neutron stars ($\delta\lesssim$300 MeV) is delimited by a horizontal solid line.

\section{Collider Prospects: FASER and FASER~2}
\label{sec:Collider}
At the LHC, the iDM states $\chi_1$ and $\chi_2$ are produced via the quark-antiquark annihilation process, i.e. the Drell–Yan process $pp \to \chi_1 \chi_2$. 
In the $M_{A'}\in [10,60]$ GeV interval considered in our work the corresponding  cross section can be approximated by the $A'$ production cross section multiplied by the branching ratio of $A' \to \chi_1 \chi_2$ (which in our case is nearly 1) in the narrow-width approximation. The production cross section scales with $\epsilon^2$ and remains almost independent of $\alpha_D$ unless $M_{A^{\prime}}\rightarrow M_Z$. 


The heavier counterpart $\chi_2$ is unstable and undergoes a three body decay to $\chi_1$ and a pair of SM particles through an off-shell $A'$, given by $\chi_2 \to \chi_1 A'* \to \chi_1 f\bar{f}$.  The expression of the relevant decay width $\Gamma_{\chi_2}$ is given in Appendix \ref{Appendix: collider}.  In particular $\Gamma_{\chi_2}$ is suppressed  at small values of the mass splitting $\delta$ so that the $\chi_2$ is a long-lived particle (LLP) on collider scales and its decay products originate from a displaced vertex spatially separated from the primary production point.  The proper decay length $c\tau$ of $\chi_2$ is given by
\begin{eqnarray}
    c\tau= \frac{1.97\times 10^{-16}}{\Gamma_{\chi_2}[\text{GeV}]} [\text{m}],
\end{eqnarray}
and the decay length in the laboratory (LHC) frame is given by $\beta \gamma c \tau$, where $\tau$ is the life time of $\chi_2$, $\beta = |\vec{p}/E|$ is the  velocity in units of the speed of light $c$ and $\gamma = E/M_{\chi_2}$ is the Lorentz factor in the laboratory frame. 
Notice that the proper decay length $c\tau$ depends only on the model parameters. On the other hand, the quantity $\beta\gamma = |\vec{p}|/M_{\chi_2}$, which characterizes the Lorentz boost of $\chi_2$, is a kinematic variable to be obtained from a numerical simulation.
 

The currently active ForwArd Search ExpeRiment (FASER) \cite{FASER:2022hcn} is well suited to detect light, weakly interacting LLPs. Such light LLPs typically have small transverse momenta ($p_T\sim M_{LLP}$, with $M_{LLP}$ the LLP mass) and are produced predominantly along the proton–proton collision axis, acquiring a large Lorentz boost in the forward direction. This can be seen in Appendix~\ref{Appendix: collider} where we provide a few examples of boost distributions along the beam axis for a set of selected benchmark points as well as the transverse boost distributions. Specifically, FASER is located 480 m downstream of the proton-proton interaction point used by the ATLAS experiment and is sensitive to LLP decays occurring inside a cylindrical detector volume with radius $R = 10$ cm and length $L = 1.5$ m. Additionally, FASER~2 has been proposed as an upgrade of FASER for the High-Luminosity LHC (HL-LHC) phase and as a part of the Forward Physics Facility (FPF) \cite{Feng:2022inv, Adhikary:2024nlv}. FASER~2 will be located  620 m downstream of the ATLAS interaction point with a cylindrical decay volume of radius $R = 1$ m and length $L = 10$ m, significantly larger than that of FASER\footnote{The decay volume of FASER~2 will likely be a $2.6\times 1\times 10$~m$^3$ cuboid \cite{FPF:2025bor}. For simplicity we assume the same cylindrical shape as in FASER.}. 


\subsection{Analysis and Results} 
We generated events consisting of Drell-Yan $p p \to \chi_1 \chi_2$ process followed by the three-body decay $\chi_2 \to \chi_1 \ell \bar{\ell}$ ($\ell = e,\mu$) using \texttt{MadGraph5\_aMC@NLO-3.5.10}~\cite{Alwall:2011uj} with \texttt{NNPDF31\_lo\_as\_0118} parton distribution function \cite{NNPDF:2017mvq}, interfaced with an \texttt{Universal FeynRules Output} (UFO) \cite{Degrande:2011ua} model from \texttt{FeynRules-2.3}~\cite{Alloul:2013bka}.
We calculated the cross sections for a 13.6 TeV center of mass energy and luminosities 300 fb$^{-1}$ and 3000 fb$^{-1}$ for FASER and FASER~2, respectively.
The probability of decay within the detector is given by 
\begin{eqnarray}
	\label{Eq: probability of decay}
	P_{\text{decay}} = e^{-L_1/{\beta \gamma c\tau}} - e^{-L_2/{\beta \gamma c\tau}}, 
\end{eqnarray}
where $L_{1,2}$ are the distances between the LHC interaction point and the points where $\chi_2$ enters and exists the detector respectively  
and the decay length in the LHC frame is given by $\beta \gamma c \tau$. 
The probability of decay inside the detector vanishes for those trajectories of $\chi_2$ which do not intersect the detector. Notice that, according to Eq.~(\ref{Eq: probability of decay}), there is always a finite probability for the $\chi_2$ to decay inside the detector, even if the characteristic proper decay length $c\tau$ falls short or exceeds the detector distance. The number of signal events in the detector volume is computed by multiplying the total number of events times the probability of decay of $\chi_2$ inside the decay volume 
\begin{eqnarray}
	\label{Eq: number of signal}
	\mathcal{N}_{\text{signal}} = \mathcal{L} \sigma P_{\text{decay}} \epsilon_{\text{det}},
\end{eqnarray}
where $\mathcal{L}$ is the luminosity, $\sigma$ is the signal cross section and $\epsilon_{\text{det}}$ is the product of the efficiency for the reconstruction of the events in the detector (which we fix $100\%$ for simplicity) and of the fraction of events passing the energy threshold for the detection of the $\chi_2$ decay products. In our analysis, following Ref.~\cite{Feng:2017uoz}, the latter is fixed by requiring the sum of the energies of the charged tracks to satisfy $E_{\text{vis}} > 100~\text{GeV}$. To compute Eq.~(\ref{Eq: number of signal}) we generated large samples of $10^6$ events for each point in the parameter space allowed by the constraints discussed in Sections~\ref{sec:collider_constraints} and~\ref{sec:early_universe_constraints}.

\begin{figure}[t]
	\centering
    \includegraphics[height=5.cm,width=7.cm]{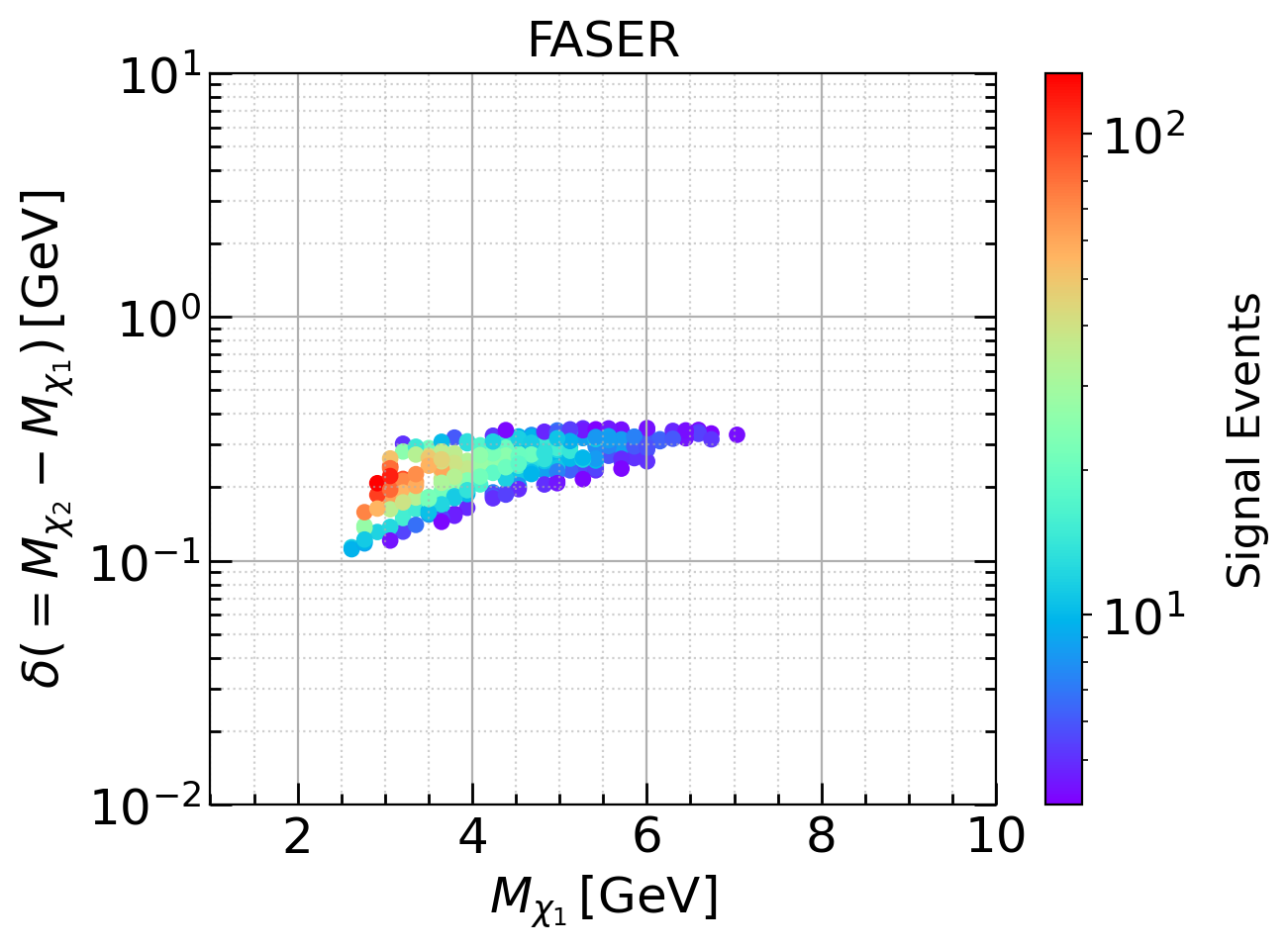}
	\includegraphics[height=5.cm,width=7.cm]{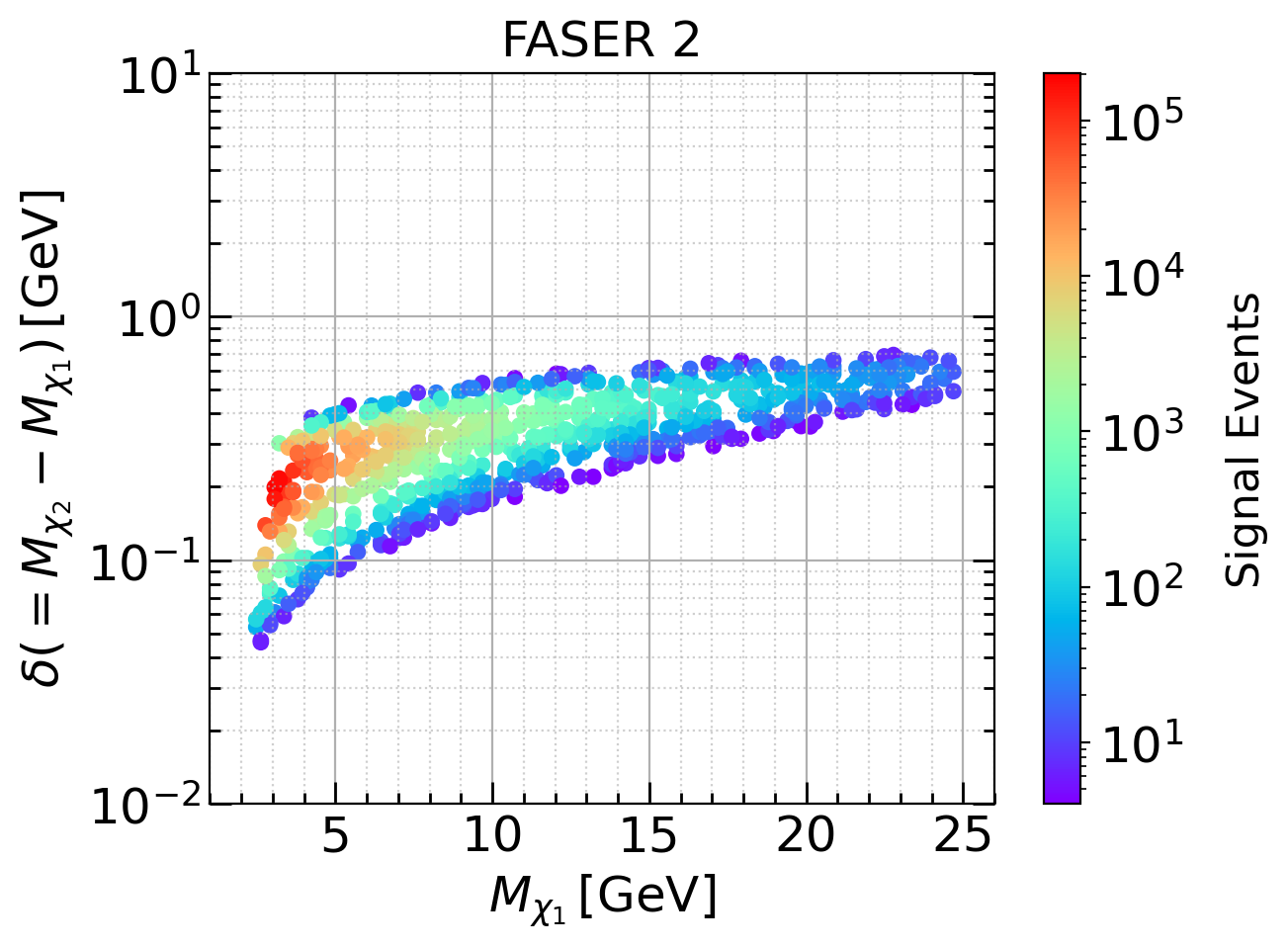}
    \includegraphics[height=5.cm,width=7.cm]{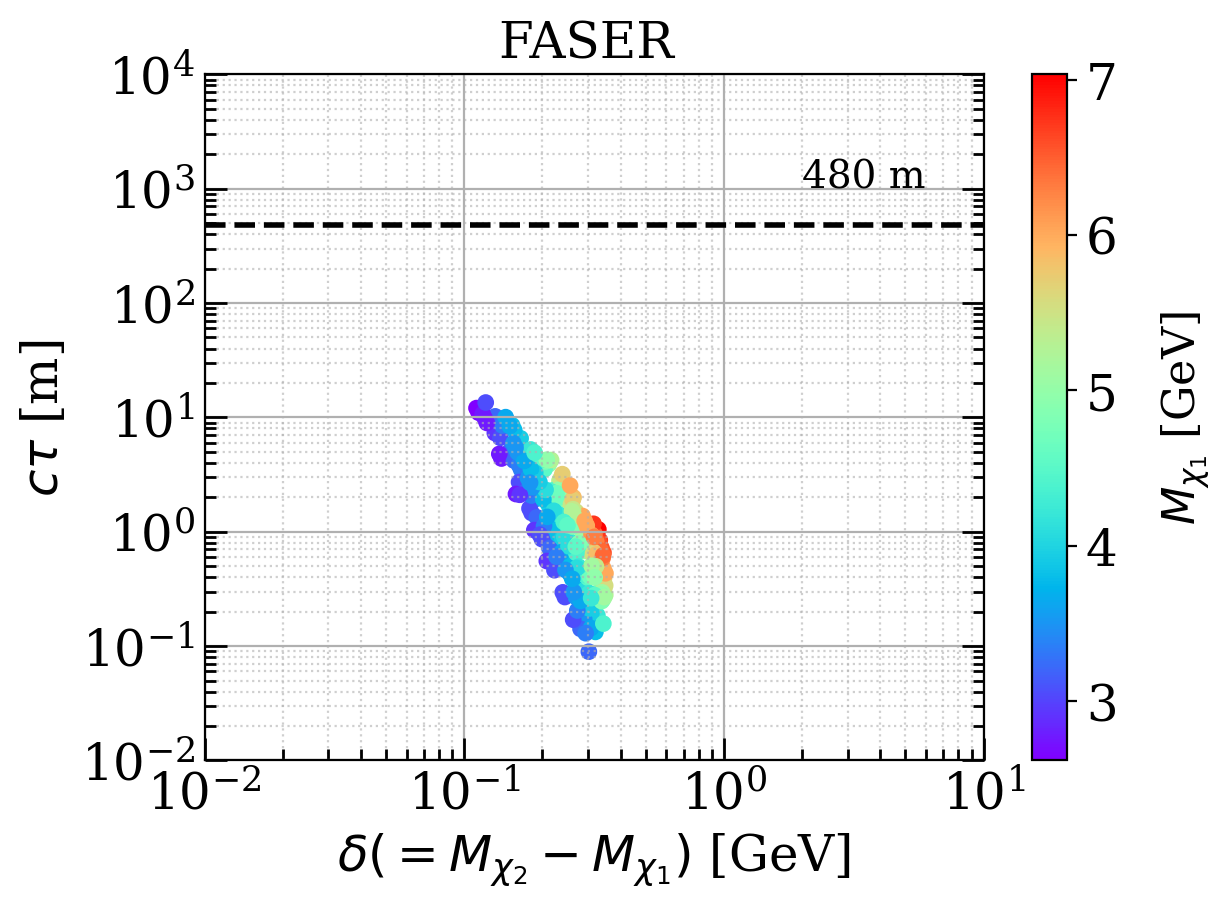}
    \includegraphics[height=5.cm,width=7.cm]{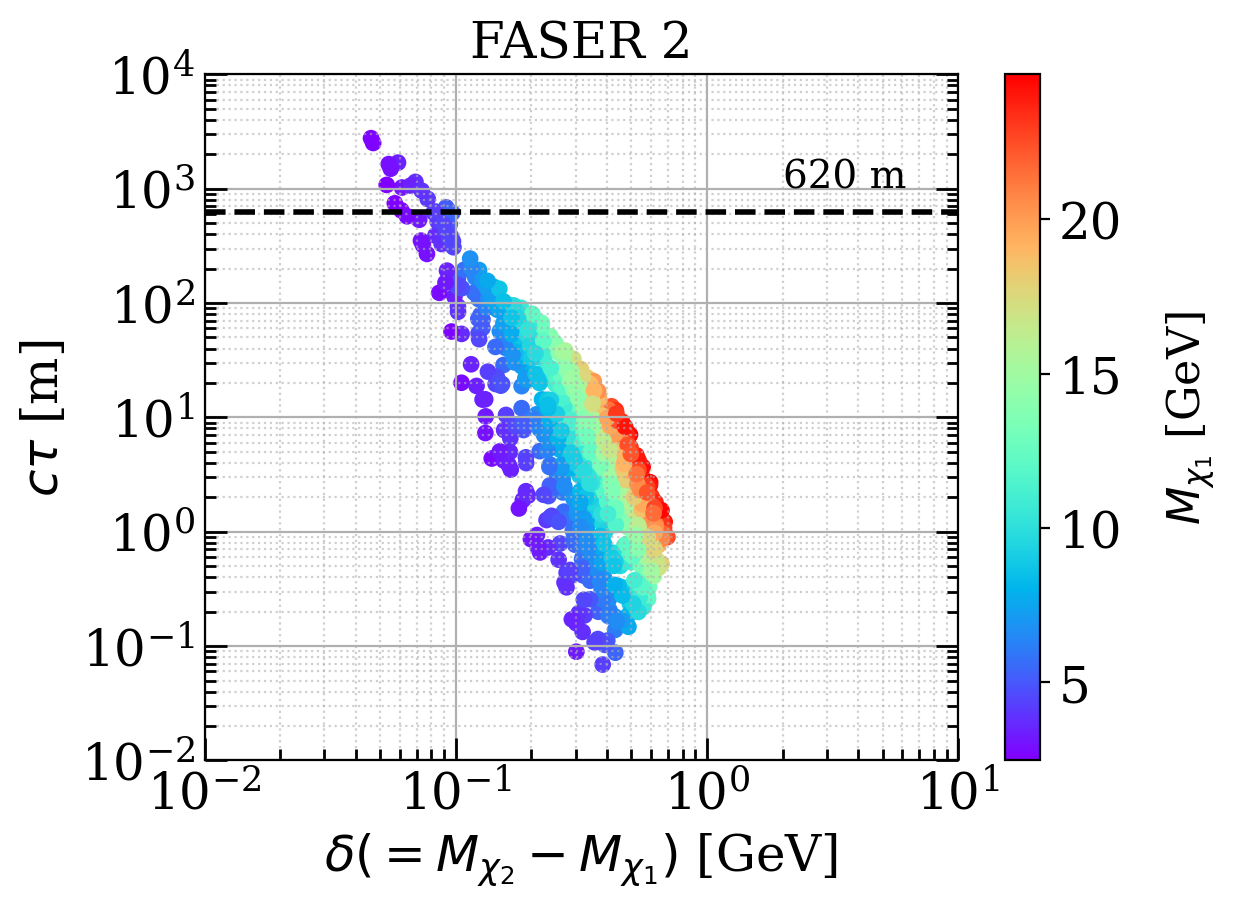}
	\caption{Scan 1, see Table~\ref{Table:Scan_parameter}. Top: Number of signal events in the FASER (left) and FASER 2 (right) decay volumes. Bottom: Range of proper decay length $c\tau$ for FASER (left) and FASER~2 (right). The horizontal dashed black lines show the distances of FASER (left) and FASER~2 (right) from the ATLAS interaction point.}
	\label{fig:Faser signal events}
\end{figure}
The results of our simulation for Scan 1 is provided in Fig.~\ref{fig:Faser signal events}. In the top panel, we show the number of signal events within the decay volumes of FASER and FASER~2. As expected, FASER~2 exhibits both a larger number of signal events and a wider sensitivity reach in the parameter space compared to FASER, owing to its larger decay volume and higher luminosity. In both experiments the number of signal events is larger when $M_{\chi_1}$ is small due to the larger $\chi_1\chi_2$ production cross section. FASER is sensitive to mass splitting values in the range 100 MeV $\lesssim \delta \lesssim 300$ MeV and to $\chi_1$ masses up to $\sim 7$ GeV and $A'$ masses up to $\sim 25$ GeV, while for FASER~2 the sensitivity is significantly extended, covering a larger parameter space with 50 MeV $\lesssim \delta \lesssim 700$ MeV and $\chi_1$ masses as large as $\simeq$ 25 GeV and $A'$ masses up to $\sim 60$ GeV. In both plots, the blue points along the boundary of the scatter plots correspond to four signal events, which defines the $2\sigma$ exclusion contours for FASER and FASER~2, assuming a negligible number of background events.  

In the lower panel of Fig.~\ref{fig:Faser signal events}, we show the range of the proper decay length $c\tau$ for FASER and FASER~2 corresponding to the scatter plots in the top panel. In particular one can notice that the Lorentz boost and the decay probability of Eq.~(\ref{Eq: probability of decay}) lead to a spread in the values of $c\tau$, which is larger in FASER 2 compared to FASER due to its larger decay volume and higher luminosity reach. In the case of FASER $c\tau$ is smaller than the detector distance, but the large longitudinal boost allows the $\chi_2$ to decay within the detector volume.  On the other hand, in FASER~2 $c\tau$ can even extend beyond the detector distance. This occurs thanks to a fraction of events with small Lorentz boosts where  at low masses of the \aidm states  the production cross sections is large enough.   

The left--hand plot of Fig.~\ref{fig:Faser signal events} shows the parameter space that FASER will probe for its projected luminosity of 300 fb$^{-1}$ (in particular, the excluded parameter space assuming no excess is observed). 


\section{Discussion and Conclusions}\label{sec:discussion_conclusions}

In this paper, we explored the phenomenology of the Dark Photon iDM ($A'\text{iDM}$) model, which is one of the simplest and most direct realizations of the iDM scenario. Our work expands upon existing analyses in three key ways: i) we extended the analysis well beyond specific benchmark points; ii) we restricted our focus exclusively to cosmologically viable configurations; iii) we evaluated the complementary relationship between accelerator and astrophysical signals.

Specifically, in Sections~\ref{sec:early_universe_constraints}, \ref{sec:neutron_stars} and \ref{sec:Collider} we examined an $A'\text{iDM}$ framework where $\alpha_D = \alpha_{EM}$ and $\epsilon = \epsilon_{max}$ (Scan 1 in Table~\ref{Table:Scan_parameter}), providing a scan of the remaining free parameters (the WIMP mass $M_{\chi_1}$, the mass splitting $\delta$, and the dark photon mass $M_{A'}$) yielding the observed relic abundance and discussing direct and indirect detection as well as potential signals from astrophysics and accelerators. 

The ensuing results  are summarized in Fig.~\ref{fig:DM_Relic_BBN}, where all the points comply with Eq.~(\ref{eqn:obs_dark_matter}), i.e. have relic densities that match observation. The cosmologically allowed parameter space corresponds to 2 GeV$\lesssim M_{\chi_1}\lesssim$25 GeV, 100 KeV $\lesssim\delta\lesssim$12 GeV and 10 GeV$\lesssim M_{A^{\prime}}\lesssim$60 GeV. In such parameter space the $\chi_1$ particle is not kinematically accessible to any target material $T$ used in DM direct searches terrestrial Direct Detection experiments, because the lowest range of $\delta$ is only possible for $m_\chi\lesssim$10 GeV, where the inelastic scattering velocity threshold $v_T^*$ of Eq.~(\ref{eq:vstar}) exceeds the maximal speed of WIMPs gravitationally bound to our Galaxy. The model is also not accessible to indirect detection from DM annihilation, because the $\chi_1\chi_2$ annihilation process is not available  since in the present Universe all the $\chi_2$ particles have decayed away, while the $\chi_1\chi_1$ annihilation process is strongly suppressed.  On the other hand we find that the projected luminosity of FASER, a detector searching for Long Lived Particles (LLP) decay at the LHC, can probe or rule out the parameters space of the model for $M_{\chi_1}\lesssim$ 7 GeV, 100 MeV  $\lesssim \delta\lesssim$ 300 MeV and $M_{A^{\prime}}\lesssim$ 25 GeV. 
The parameter space accessible to the LLP search would extend to the full range of $m_{\chi_1}$ for the FASER 2 upgrade of FASER. Moreover, the relic abundance constraint implies that in the same scan the WIMP--neutron cross section exceeds the geometrical cross section of a neutron star with mass 1.5 $M_\odot$ and R=10 km. In this regime all the $\chi_1$ particles that are on the neutron star's path are captured and deposit almost all their kinetic energy in the celestial body through a large number of scattering events. This process is expected to heat--up the neutron star close to the Earth to $\simeq$ 2000 Kelvin if $\delta\lesssim$300 MeV (the maximal mass splitting kinematically accessible to the gravitational acceleration of the neutron star). This may provide a potential, albeit challenging, way to detect the Dark Photon iDM scenario in a region of parameter space that partially overlaps with LLP searches at the LHC.

The results and conclusions discussed so far for Scan 1 are not modified substantially when also $\alpha_D$ and $\epsilon$ are allowed to vary (5--dimensional Scan 2 in Table~\ref{Table:Scan_parameter}). This is discussed in detail in Appendix~\ref{app:5dim} (see the matrix plot of Fig.~\ref{fig:Matrix_Plot_Scan 2}) where rather than plotting $\alpha_D$ and $\epsilon$ separately we combine them showing instead the phenomenologically sensible combination $\zeta=\epsilon^2\alpha_D$. In practice, the only noteworthy change in such plots is that additional cosmologically allowed configurations are present at lower values of $m_{\chi_1}$, extending the range of Scan 1 down to $m_{\chi_1}=1$ GeV. For the same configurations the cosmologically allowed range for the $\delta$ parameter slightly lowers a few tens of keV. In particular, our conclusions for Scan 1 still hold for Scan 2: the model remains kinematically beyond the reach of terrestrial DD experiments,  and the parameter space accessible via
accelerator LLP searches partially overlaps with the region probed by $\chi_1$ capture in neutron stars (that always saturates the optically thick limit as in Fig~\ref{fig:DM_NS} for Scan 1).  Such mild sensitivity to the variation of the $(\alpha_D, \epsilon)$ parameters can be understood with the strong correlation through the same $\zeta/M_{A^{\prime}}^4$ dependence among $\langle\sigma v\rangle_{eff}$, $\sigma_{el}$ and $\Gamma(\chi_2 \to \chi_1 \ell\bar{\ell})$, combined with the steep dependence of the latter on the $\delta$ parameter ($\Gamma(\chi_2 \to \chi_1 \ell\bar{\ell})\propto \delta^5$, see Eq.~(\ref{Eq: chi_2 to leptons})): once $\langle\sigma v\rangle_{eff}$ is pinned to the observed relic density by Eq.~(\ref{eqn:obs_dark_matter}) also the other quantities fall into line. Moreover, in the optically thick regime capture in neutron stars remains insensitive to the microphysics, and $\delta$ can only change by a small amount without driving the $\chi_2$ decay length outside the reach of FASER and FASER 2.


\section*{Acknowledgments}
This research was supported by Basic Science Research Program through the National Research Foundation of Korea (NRF) funded by the Ministry of Education through the Center for Quantum Spacetime (CQUeST) of Sogang University with grant number RS-2020-NR049598 and by the Ministry of Science and ICT with grant number RS-2025-24523022. PS would like to thank Felix Kling for useful discussions. PS also acknowledges the KIAS Center for Advanced Computation for providing computing resources. AR would also like to thank Rojalin Padhan for useful discussion.


\appendix
 
\section{Full (five--dimensional) scan}
		\label{app:5dim}
    The analysis of Sections~\ref{sec:early_universe_constraints}, \ref{sec:neutron_stars} and \ref{sec:Collider} was performed using a 3--dimensional scan of the \aidm parameter space (Scan 1 in Table~\ref{Table:Scan_parameter}) where the kinetic mixing parameter was fixed to $\epsilon = \epsilon_{MAX}$ and the dark-sector coupling to $\alpha_D = \alpha_{\mathrm{EM}}$. Such choice allows a particularly transparent discussion with two--dimensional plots with a color-code representing the third dimension.
    
    In this appendix we wish to extend the scan to the full 5 dimensions of the \aidm model. The bottom-line of this Section is that, within the parameter space yielding a correct relic abundance, the \aidm phenomenology changes little between a 3-dim plot and a 5-dim scan, besides the fact that additional configurations at low $m_\chi$ ($m_\chi\lesssim$ 1 GeV) are obtained, that turn out to be detectable both in NS and in Faser.   
        In particular, our conclusions that the sensitivities to neutron-stars and FASER (i.e. between astrophysics and colliders) overlap extend seamlessly to the 5-dimensional scan, i.e. they do not require any tuning of the $\alpha_{D}$ and $\epsilon$ parameters. 
        
        The range of parameters used in the five-dimensional scan (Scan 2) is given in Table I. Specifically, we allowed for 
	    larger values of the dark gauge coupling (bounded by perturbativity), and for lower values of $\epsilon$ (saturating to the overclosure limit where a correct relic abundance is no longer possible).  
	
\begin{figure}[t]
	\centering
\includegraphics[height=12.cm,width=14.cm]{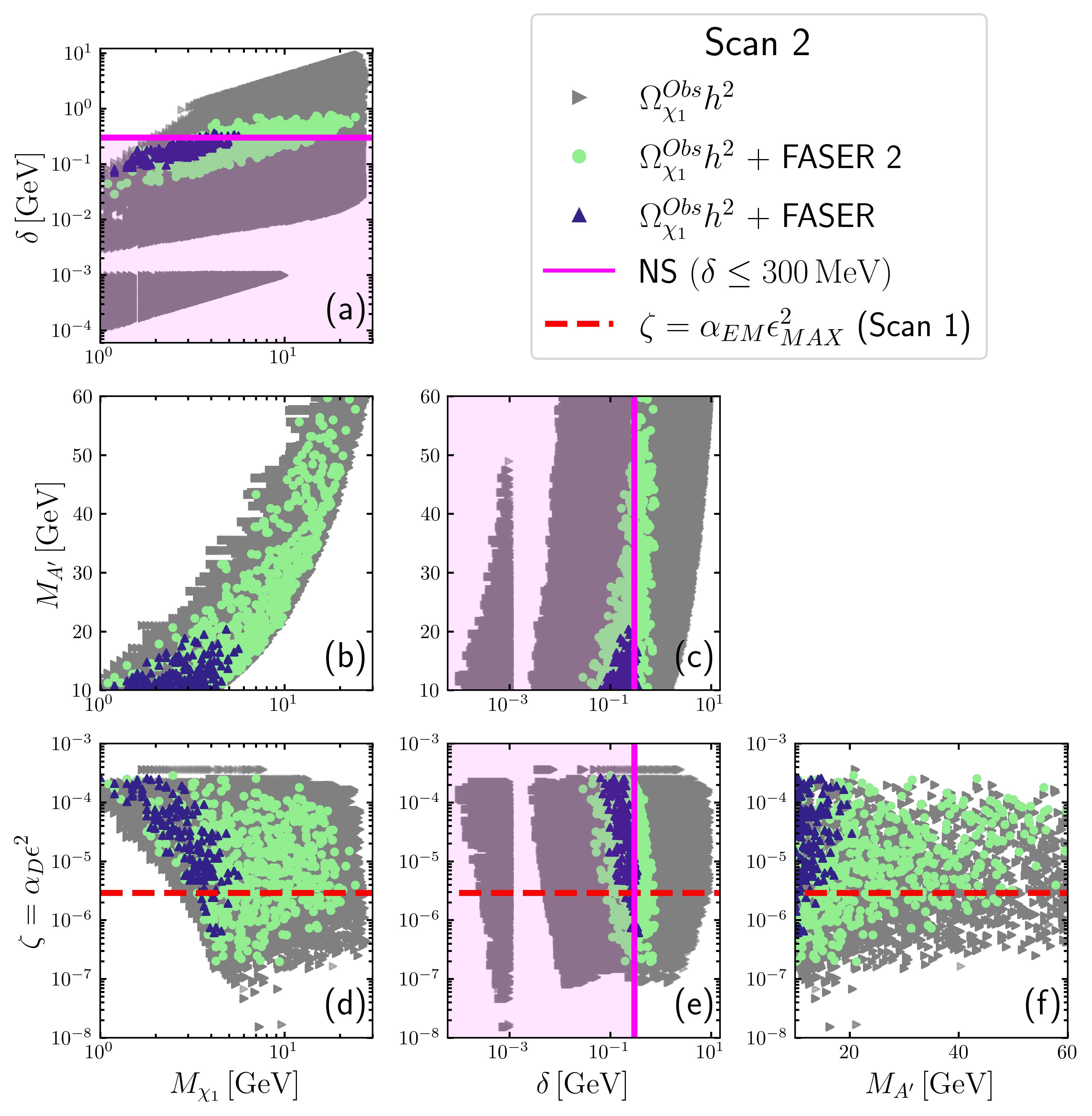}
	\caption{Allowed parameter space Scan 2, see Table~\ref{Table:Scan_parameter}. }
	\label{fig:Matrix_Plot_Scan 2}
\end{figure}

	The results of the five-dimensional scan are shown in Fig.~\ref{fig:Matrix_Plot_Scan 2}. The grey points satisfy the observed relic density, while the blue and green overlayed points indicate the parameter space within the sensitivity reach of FASER and FASER 2, respectively. Such plots need to be compounded by Fig.~\ref{fig:DM_NS}, that shows that in all the configurations of Scan 2 capture in neutron stars saturates the geometric limit. 
	
	Specifically, Fig.~\ref{fig:Matrix_Plot_Scan 2} subplot (a) shows that the overlap between the sensitivity reaches of FASER, FASER 2 and neutron stars is nearly identical to that found in Scan 1. Along with that, we find additional cosmologically allowed parameter space at lower values of $m_{\chi_1}$, extending down to $m_{\chi_1}=1$ GeV. In Scan 2 the reach of FASER is $1~\text{GeV} \lesssim m_{\chi_1} \lesssim 7~\text{GeV}$, $60~\text{MeV} \lesssim \delta \lesssim 400~\text{MeV}$ and the reach of FASER 2 is $1~\text{GeV} \lesssim m_{\chi_1} \lesssim 25~\text{GeV}$, $30~\text{MeV} \lesssim \delta \lesssim 800~\text{MeV}$. The pink shaded region at $\delta \lesssim 300$ MeV indicates the neutron-star sensitivity reach. 
	In subplot (b), we show the correlation between $M_{A'}$ and $M_{\chi_1}$, demonstrating that a substantial number of parameter points satisfy the relic-density constraint even when $\langle \sigma v \rangle_{eff}$ is evaluated away from the resonance. In particular, we notice that the range 10 GeV$\le m_{A'}\le$ 60 GeV that we adopted to avoid the BaBar constraint~\cite{BaBar:2017tiz,BaBar:2008aby} and to avoid the 
    precision studies at LEP~\cite{Hook:2010tw,Curtin:2014cca} and displaced muon jet searches at CMS~\cite{CMS:2023bay} were already saturated in Scan 1, so they remain identical in Scan 2. 

	For subplots (d), (e), and (f), we define the combination $\zeta = \alpha_D\epsilon^2$, that enters in $\langle \sigma v \rangle_{\mathrm{eff}}$ (relic abundance), in $\sigma_{\text{inel}}$ (capture in neutron stars) and in the decay length of $\chi_2$ (FASER) (it is worth pointing out that two quantities do not depend on the $\zeta$ combination: the decay width of $A'$,which depends only on $\alpha_D$, and the Drell–Yan $\chi_1 \chi_2$ production cross section, which depends on $\epsilon^2$). 
    
    The red dashed horizontal line corresponds to $\zeta = \alpha_{\mathrm{EM}} \epsilon_{MAX}^2$, the value used in Scan 1. Subplot (d) shows that the additional parameter space at lower values of $M_{\chi_1}$ allowed by the relic density constraint  comes from larger values of $\zeta$ (i.e. $\alpha_D$), which enhance $\langle \sigma v \rangle_{\mathrm{eff}}$. These new configurations are detectable in FASER, FASER 2 and in neutron stars. 
    
    Subplot (e) shows that the variation of $\delta$ with respect to $\zeta$ is marginal. Indeed, the range of $\delta$ is mostly driven by the relic abundance (that has exponential sensitivity on $\delta$ through $\langle \sigma v \rangle_{\mathrm{eff}}$,  see Eq.~(\ref{eqn:DM_SigmaV})), and by FASER's baseline (which depend on $\delta^5$ through the $\chi_2$ decay width, see Eq.(\ref{Eq: chi_2 to leptons})).  Moreover, subplot (e) also shows that the range of $\zeta$ where a correct relic abundance is achieved is almost saturated\footnote{For both FASER and FASER 2, we find a lower and upper thresholds in $\zeta$ beyond which sensitivities are lost. Decreasing (increasing) $\zeta$ increases (decreases) the decay length of $\chi_2$, reducing the probability that it decays within the detector volume. The lower threshold in $\zeta$ is more restrictive for FASER, since decreasing $\epsilon$ also reduces the $\chi_1\chi_2$ production cross section. This reduction has also a greater impact on FASER because of its smaller decay volume and lower luminosity.} by that where both FASER and FASER 2 are observable while, at the same time, a large part of the same configurations lie below the kinematic threshold for capture in neutron stars (represented by the vertical solid line). In other words, as already pointed out the possibility of probing \aidm both through astrophysical observations and accelerator experiments does not rely on a fine-tuning of the $\alpha_D$ and $\epsilon$ parameters.



\section{Decay widths and Lorentz boost}
\label{Appendix: collider}
For $M_{A'} > M_{\chi_1} + M_{\chi_2}$, $A'$ decays to $\chi_1 \chi_2$ by almost $100\%$ and the decay width is given by \cite{Foguel:2024lca}
\begin{eqnarray}
	\label{Eq:A' decay to chi_1 chi_2}
	\Gamma (A' \to \chi_1\chi_2) &=& \frac{\alpha_D}{3}M_{A'}\left ( 1-\frac{\delta^2}{M_{A'}^2} \right)^{3/2}\left(1 + \frac{( M_{\chi_1} + M_{\chi_2})^2}{2 M_{A'}^2}\right) \nonumber \\
	&\times&\sqrt{1-\frac{(M_{\chi_1} + M_{\chi_2})^2}{M_{A'}^2}}.
\end{eqnarray}
The three body decay width of $\chi_2$ to a pair of leptons in the limit $M_{A'} \gg M_{\chi_2} > M_{\chi_1}$ is given by \cite{Izaguirre:2015zva, Berlin:2018jbm}
\begin{eqnarray}
	\label{Eq: chi_2 to leptons}
	\Gamma(\chi_2 \to \chi_1 \ell\bar{\ell}) = \frac{4 \epsilon^2 \alpha_{EM}\alpha_D \delta^5}{15 \pi M_{A'}^4}. 
\end{eqnarray}
The perturbative computation for the hadronic decay mode of $\chi_2$ applies only when $\delta \gtrsim 2$ GeV.
Therefore, the total decay width of $\chi_2$ is obtained from the $\chi_2 \to \chi_1 e^+ e^-$ process by including the branching ratio (BR) of an off-shell $A'^*$ decaying into $e^+e^-$ pair evaluated at the invariant mass $M^{\text{inv}}_{ee}$. Following Refs.~\cite{Giudice:2017zke, Jodlowski:2019ycu}, $\Gamma_{\chi_2}$ is given by
\begin{eqnarray}
	\label{Eq: chi_2 total decay}
	\Gamma_{\chi_2} &=& \frac{\epsilon^2\alpha_D \alpha_{EM}}{16\pi M_{\chi_2}^3}\int_{s_2^-}^{s_2^+} ds_2 \int_{s_1^-}^{s_1^+} ds_1 \frac{4|A|^2}{(M_{\chi_1}^2 + M_{\chi_2}^2 + 2 M_e^2 -s_1 - s_2 -M^2_{A'})^2 + M_{A'}^2\Gamma_{A'}^2} \nonumber \\
	&&\times \frac{1}{\text{BR}(M_{A'^*}=M^{\text{inv}}_{ee})},
\end{eqnarray}
where $\Gamma_{A'}$ is the total decay width of $A'$ and is given in Eq.~(\ref{Eq:A' decay to chi_1 chi_2}) and 
\begin{eqnarray}
	|A|^2 &=& ( s_1 + s_2  - 2 M_{\chi_1}M_{\chi_2} - 2M_e^2)[(M_{\chi_1}+ M_{\chi_2})^2 + 4 M_e^2] +2(M_e^2 + M_{\chi_1}M_{\chi_2})^2 \nonumber \\
	&&- s_1^2 -s_2^2.
\end{eqnarray}
The integral limits are given by 
\begin{eqnarray}
	\label{Eq: Integral limits}
	s_1^\pm &=& M_{\chi_1}^2 + M_{e}^2 + \frac{1}{2s_2}\Big[ (M^2_{\chi_2} - M^2_e - s_2)(M^2_{\chi_1} - M^2_e + s_2) \nonumber \\
	&&\pm \lambda^{1/2}(s_2, M^2_{\chi_2}, M^2_e) \lambda^{1/2}(s_2, M^2_{\chi_1}, M^2_e) \Big],  \nonumber \\
	s_2^+ &=& (M_{\chi_2} - M_e)^2  \quad \text{and} \quad  s_2^- = (M_{\chi_1} + M_e)^2,
\end{eqnarray}
where $\lambda(x,y,z) = (x-y-z)^2 - 4 yz$. To evaluate $\text{BR}(M_{A'^*}=M^{\text{inv}}_{ee})$ in Eq.~\ref{Eq: chi_2 total decay}, we considered the decay widths of $A'$  into SM fermions
\begin{eqnarray}
	\Gamma(A' \to f\bar{f}) =  N_C \frac{Q_f^2 \epsilon^2 \alpha_{EM}}{3} M_{A'} \left( 1 + 2 \frac{M^2_f}{M_{A'}^2} \right)\sqrt{1-\frac{4M_f^2}{M_{A'}^2}},
\end{eqnarray}
where $N_C$ = 3 (1) is the color charge of the quarks (leptons) and $Q_f$ is the electric charge. For $M_{A^{\prime}} \lesssim 2$ GeV, the above perturbative calculation of the $A' \to q\bar{q}$ decay width fails~\cite{Graham:2021ggy,Foguel:2022ppx}. However, in this case the latter can be obtained from the experimentally measured value of $R_\mu^{\mathcal{H}} = \sigma(e^+ e^- \to \mathcal{H})/\sigma(e^+ e^- \to \mu^+ \mu^-)$ \cite{ParticleDataGroup:2024cfk}, such that   
\begin{eqnarray}
	\Gamma(A' \to \mathcal{H}) = \Gamma(A' \to \mu^+ \mu^-)R_\mu^{\mathcal{H}}(M^2_{A'}),
\end{eqnarray}
which automatically accounts for the mixing with the QCD vector mesons, the $\rho$, $\omega$ and $\phi$, along with all other non-perturbative QCD effects.

\begin{figure}[t]
	\centering
	\includegraphics[height=5.cm,width=7.cm]{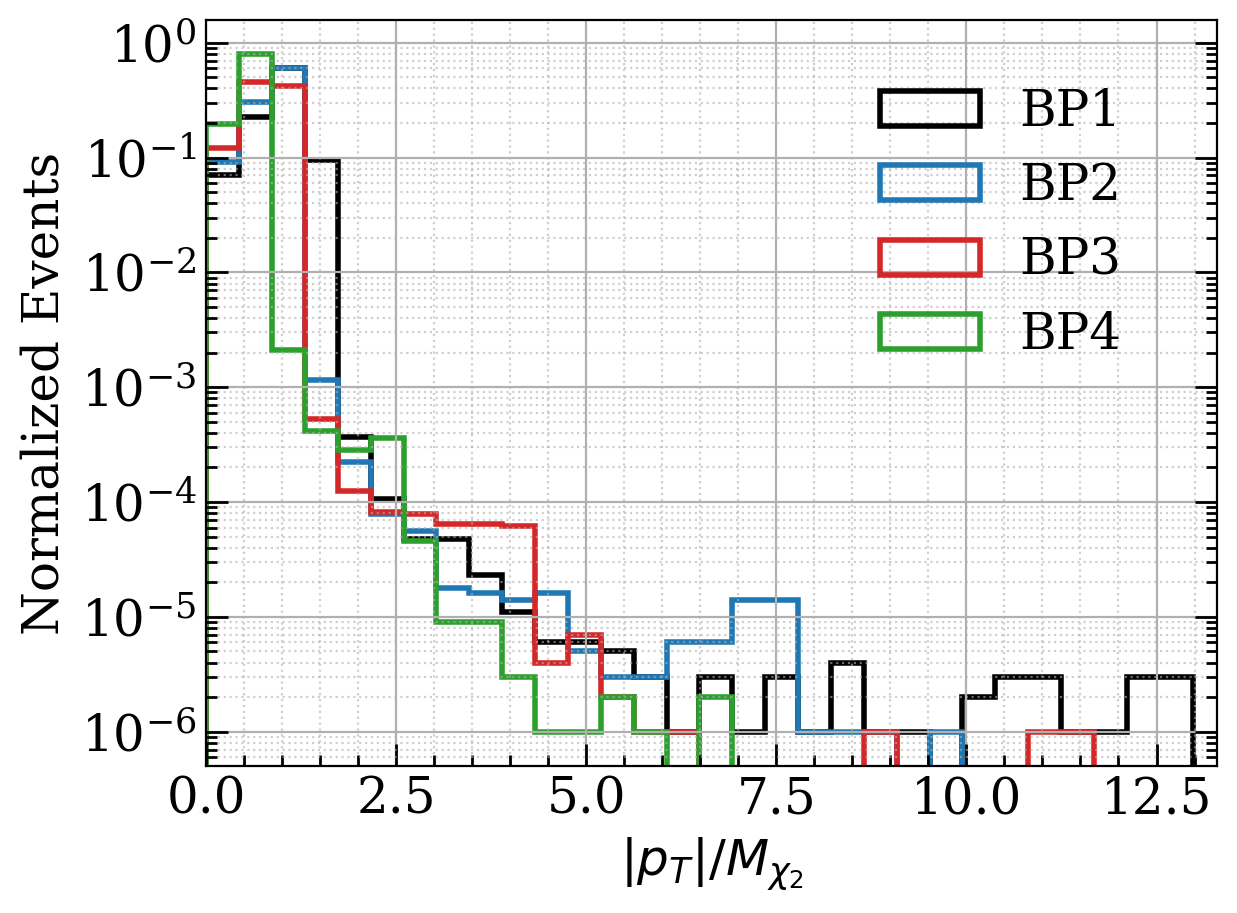}
	\includegraphics[height=5.cm,width=7.cm]{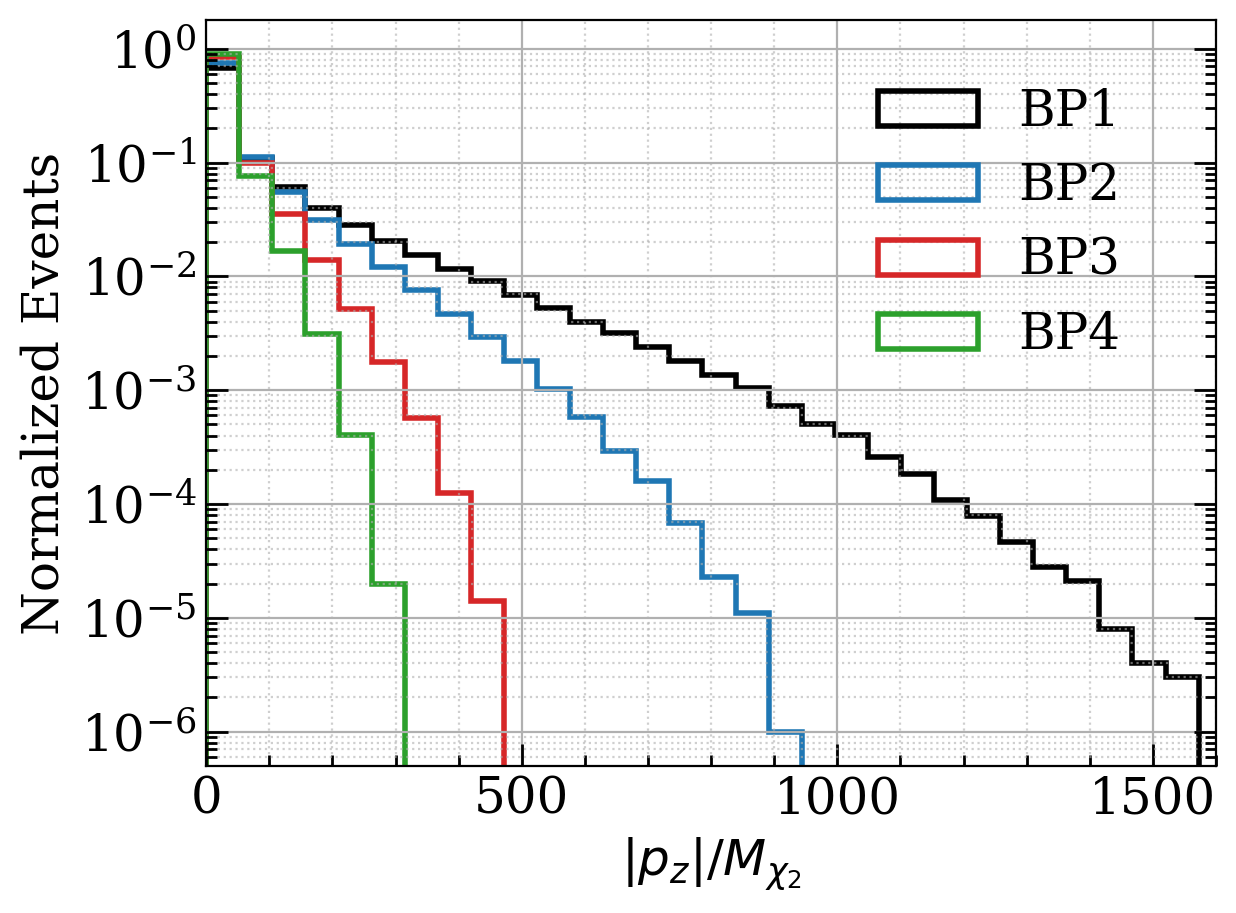}
	\caption{Normalized transverse (left) and longitudinal (right)  boost distributions for the benchmark points in Eq.~(\ref{benchmark points 1}).}
	\label{fig:Boost}
\end{figure}

In Eq.~(\ref{Eq: probability of decay}), $\beta\gamma c\tau$ denotes the decay length of $\chi_2$ in the laboratory or the LHC frame, where $\beta\gamma = |\vec{p}|/M_{\chi_2}$ represents the Lorentz boost, which is a purely kinematic quantity. In Fig.~\ref{fig:Boost}, we provide examples of the transverse and longitudinal boost distributions for the following benchmark points:
\begin{eqnarray}
	\label{benchmark points 1}
	\text{BP1:}&& M_{\chi_1} = 3.35 ~\text{GeV}, ~~\delta= 0.21 ~\text{GeV}, ~~ M_{A'} = 11.60 ~\text{GeV}, \nonumber \\
	\text{BP2:}&& M_{\chi_1} = 5.56 ~\text{GeV}, ~~\delta= 0.32 ~\text{GeV} , ~~ M_{A'} = 17.62 ~\text{GeV}, \nonumber \\
	\text{BP3:}&& M_{\chi_1} = 10.71 ~\text{GeV}, ~~\delta= 0.27 ~\text{GeV}, ~~M_{A'} = 30.45 ~\text{GeV}, \nonumber \\ 
	\text{BP4:}&& M_{\chi_1} = 16.15 ~\text{GeV}, ~~\delta= 0.56 ~\text{GeV}, ~~M_{A'} = 41.68 ~\text{GeV}.
\end{eqnarray}
where we fix $\alpha_D = \alpha_{EM}$ and $\epsilon = \epsilon_{max}(M_{A^{\prime}})$.
The corresponding values for the proper decay length $c\tau$, the production cross section $\sigma$ of the process $pp \to \chi_1 \chi_2, \chi_2 \to \chi_1 \ell\ell$ and the number of signal events in FASER (FASER~2) are given by:
\begin{eqnarray}
	\label{benchmark points 2}
	\text{BP1:}&& c\tau = 0.94 ~\text{m}, ~~\sigma = 1332 ~\text{pb}, ~~\text{signal events} = 64 ~(138474), \nonumber \\ 
	\text{BP2:}&& c\tau = 0.57 ~\text{m}, ~~\sigma = 549 ~\text{pb},  ~~\text{signal events} = 9 ~(14643), \nonumber \\
	\text{BP3:}&& c\tau = 11.32 ~\text{m}, ~~\sigma = 139 ~\text{pb},  ~~\text{signal events} = 0 ~(228), \nonumber \\ 
	\text{BP4:}&& c\tau = 0.86 ~\text{m}, ~~\sigma = 75 ~\text{pb},  ~~\text{signal events} = 0 ~(69).
\end{eqnarray}
Fig.~\ref{fig:Boost} shows that within the parameter space of interest, the longitudinal or forward boost (right) is significantly larger than the transverse boost (left). It also indicates that the forward boost becomes more pronounced for lower masses of $\chi_2$.

\bibliographystyle{JHEP}
\bibliography{bibitem.bib}
\end{document}